\documentclass[prb,superscriptaddress,twocolumn,showkeys]{revtex4}
\usepackage{graphicx}
\usepackage{hyperref}
\usepackage{amsmath}
\usepackage{amssymb}
\usepackage{bm}
\usepackage{units}
\usepackage{xcolor}
\usepackage{ulem}

\newcommand{\lcpq}{Laboratoire de Chimie et Physique Quantiques, Universit\'e de Toulouse, UPS, CNRS, and European Theoretical Spectroscopy Facility (ETSF), 118 route de Narbonne, F-31062 Toulouse, France}
\newcommand{\lpt}{Laboratoire de Physique Th\'eorique, CNRS, Universit\'e de Toulouse, UPS, and European Theoretical Spectroscopy Facility (ETSF), 118 route de Narbonne, F-31062 Toulouse, France}
\newcommand{\wisconsin}{Department of Materials Science and Engineering, University of Wisconsin-Madison, 53706, USA}

\begin{document}

\title{Multichannel Dyson equation: derivation and analysis}

\author{Gabriele Riva}
\affiliation{\wisconsin}
\email{griva@wisc.edu}
\author{Pina Romaniello}
\affiliation{\lpt}
\email{pina.romaniello@irsamc.ups-tlse.fr}
\author{J. Arjan Berger}
\email{arjan.berger@irsamc.ups-tlse.fr}
\affiliation{\lcpq}

\begin{abstract}
In a recent letter [Phys.~Rev.~Lett. 131, 216401] we presented the multichannel Dyson equation (MCDE) in which two or more many-body Green's functions are coupled.
In this work we will give further details of the MCDE approach.
In particular we will discuss: 1) the derivation of the MCDE and the definition of the space in which it is to be solved; 2) the rationale of the approximation to the multichannel self-energy; 3) a diagrammatic analysis of the MCDE; 4) the recasting of the MCDE on an eigenvalue problem with an effective Hamiltonian that can be solved using standard numerical techniques.
This work mainly focuses on the coupling between the one-body Green's function and the three-body Green's function to describe photoemission spectra, but the MCDE method can be generalized to the coupling of other many-body Green's functions and to other spectroscopies. 
\end{abstract}

\maketitle

\section{Introduction}

We have recently presented the multichannel Dyson equation (MCDE) as an alternative to the standard single-channel Dyson equation.~\cite{Riv23}
While the latter involves a single many-body Green's function, the MCDE uses a multichannel Green's function in which two or more many-body Green's functions are coupled.
To demonstrate the advantages of the MCDE we have mainly focused on the coupling of the one-body Green's function (1-GF) and the three-body Green's function (3-GF), more precisely its two-hole-one-electron and two-electron-one-hole parts, to calculate direct and inverse photoemission spectra.~\cite{Riv22,Riv23}
We showed that this coupling naturally puts quasiparticles and satellites on an equal footing contrary to the single-channel Dyson equation for the 1-GF.
We illustrated this principle by applying the MCDE to the Hubbard dimer at 1/4 and 1/2 filling for which we obtained the exact spectral functions for all interaction strengths. Alternative approaches based on the single-channel Dyson equation for the 1-GF such as $GW$ are not able to yield exact results. 
In particular satellites are not well described with these methods, especially at large interaction strengths.~\cite{Dis21-2,Dis23-1}

Besides this improved description of satellites in photoemission spectra the MCDE has several other advantages: 1) the multichannel self-energy is not a functional of the multichannel Green's function and only involves the bare Coulomb potential.~\footnote{The evaluation of the MCDE requires Hartree-Fock orbitals and energies which are obtained from a self-consistent procedure.}
As a consequence, no self-consistency is required to solve the MCDE. Moreover, one thus avoids the problem to run into an unphysical solution.~\cite{Lan12,Ber14,Sta15,Tar17}
2) the multichannel self-energy is static and Hermitian. Therefore, the MCDE can be rewritten as an eigenvalue problem with an effective Hamiltonian.
There exist standard numerical techniques to solve such an equation.~\cite{Hay72,Schm03,Her05}

We note that an alternative approach to calculate the 1-GF using a configuration space that includes particle-hole excitations is the adiabatic diagrammatic construction~\cite{Sch78,Sch83,Nie84,Bin21} which takes its origin in approaches developed in nuclear physics~\cite{Ethofer1969,Schuck_1973,Sch04,Bar01,Bar07} that use the 3-GF as the fundamental quantity. So far the 3-GF has not been used much in electronic-structure calculations although there are some notable exceptions~\cite{Man94,Mar99,Dei16,Tor19,Pav21,Kapple_2023}.

Here we will give further details of the derivation of the MCDE and analyse its properties.
We will mainly focus on the rationale behind the approximation to the multichannel self-energy.
In our approximation of the multichannel self-energy we include all first-order interactions.
As a consequence there are only interactions between pairs of particles, i.e., two electrons, two holes, or an electron and a hole.
Therefore, we can use the kernels that have been developed for the elecron-hole and particle-particle Bethe-Salpeter equation (BSE) to build the multichannel self-energy.
~\cite{Oni02,Rom12} We note that, although the multichannel self-energy only involves first-order interactions, the multichannel Green's function obtained by solving the MCDE will contain interactions to all orders.
In this work we will also give more details about the diagrammatic analysis of the MCDE.
Finally, we will also address in detail how the MCDE can be recast as an eigenvalue problem with an effective Hamiltonian.

The paper is organized as follows.
In section \ref{sec:MCDE} we derive the MCDE and in section \ref{sec:MCSE} we analyse the kernels of the electron-hole and particle-particle BSE and use it to construct the kernel of the MCDE.
We perform a diagrammatic analysis of the MCDE in section \ref{sec:analysis} and show how it can be solved in practice in section \ref{sec:Hamiltonian} by rewriting the MCDE in terms of an eigenvalue equation with an effective Hamiltonian.
Finally, we draw our conclusions in section. \ref{sec:conclusions}

\section{The multichannel Dyson equation}
\label{sec:MCDE}

\subsection{The three-body Green's function}

Our starting point is the time-ordered equilibrium 3-GF at zero temperature, which is defined by
\begin{align}\label{G3def:eq}
    &G_3(1,2,3,1',2',3')=\nonumber\\
    &i\langle \Psi_0^N|\hat{T}[\hat{\psi}(1)\hat{\psi}(2)\hat{\psi}(3) \hat{\psi}^{\dagger}(3')\hat{\psi}^{\dagger}(2')\hat{\psi}^{\dagger}(1') ]|\Psi_0^N\rangle,
\end{align}
where $\hat{\psi}$ and $\hat{\psi}^{\dagger}$ are the creation and annihilation operators, respectively, $|\Psi_0^N\rangle$ is the $N$-electron ground-state and $\hat{T}$ is Wick's time-ordering operator. 
We used the notation $1=(x_1,t_1)$, with $x_1$ referring to both position and spin. 
The 3-GF describes the propagation of three particles, namely, three electrons ($3e$), three holes ($3h$), two electrons and one hole ($2e1h$), or two holes and one electron ($2h1e$). In this work, we want to describe direct photoemission spectroscopy (PES) and inverse photoemission spectroscopy (IPES). 
Therefore, we focus only on the last two cases, which describe the simultaneous propagation of a particle (electron or hole) and an electron-hole ($eh$) pair. In general, the 3-GF depends on five time differences when the Hamiltonian is time-independent. However, we are interested only in the time difference that describes the combined propagation of one particle and an $eh$ pair. All the other time differences describe phenomena that can be considered instantaneous, such as the time between the addition of an electron and the creation of the $eh$ pair or the time to create the $eh$ pair~\cite{Riv22}.  

Taking into account the above considerations, the part of the 3-GF that is of interest to describe PES and IPES is given by
\begin{align}
    &G_3(x_1,x_2,x_3,x_{1'},x_{2'},x_{3'};t-t')=-i \,\, \times\nonumber \\
    &\langle\! \Psi_0^N \!| \hat{T}[(\hat{\psi}^{\dagger}\!(x_{3'}\!)\hat{\psi}(x_2\!)\hat{\psi}(x_1\!))_{t} (\hat{\psi}^{\dagger}\!(x_{1'})\hat{\psi}^{\dagger}\!(x_{2'})\hat{\psi}(x_3))_{t'}] |\!\Psi_0^N\! \rangle
\end{align}
where the notation $(\hat{\psi}^{\dagger}(x_{3'})\hat{\psi}(x_2)\hat{\psi}(x_1))_{t}$ implies that the three field operators act at the same time $t$. 
We write the following Dyson equation for $G_3$ in frequency space~\cite{Riv22}
\begin{equation}\label{Dyson:eq}
    G_{3}(\omega)=G_{3}^0(\omega)+G_{3}^0(\omega) \Sigma_{3}(\omega)G_{3}(\omega),
\end{equation}
where $G^0_{3}(\omega)$ is the noninteracting 3-GF and $\Sigma_3(\omega)$ is the three-body self-energy, which is defined by Eq.~\eqref{Dyson:eq}. 
We project this equation in the basis set of one-electron spinorbitals $\{\phi_i\}$ that diagonalizes $G^0_3$. By expressing the field operators in this basis as
\begin{align}\label{projection:eq}
    \hat{\psi}(x)=\sum_p \phi_p(x) \hat c_p & &     \hat{\psi}^{\dagger}(x)=\sum_p \phi^{*}_p(x) \hat c^{\dagger}_p,
\end{align}
the 3-GF becomes
\begin{align}\label{G3changebasis:eq}
    &G_{ijl;mok}(\omega)=\int dx_1dx_2dx_3dx_{1'}dx_{2'}dx_{3'} \phi_i^*(x_1)\phi_j^*(x_2)\nonumber\\
    &\phi_l(x_{3'}\!)\phi_m(x_{1'}\!)\phi_o(x_{2'}\!)\phi_k^*(x_3)G(x_1,x_2,x_3,x_{1'},x_{2'},x_{3'};\omega).
\end{align}
Here and in the following we will suppress the subscript of an $n$-body Green's function $G_n$ whenever the number of arguments or indices of the Green function are explicitly given, e.g., $G_{3\,ijl;mok} \to G_{ijl;mok}$.

Its spectral representation is given by~\cite{Riv22}
\begin{align}\label{G3basis:eq}
    G_{ijl;mok}(\omega)&=\sum_n \frac{\textrm{X}_n^{ijl}\textrm{X}_n^{\dagger \; mok}}{\omega-(E_n^{N+1}-E_0^N)+i\eta}\nonumber \\
    &+\sum_n \frac{\textrm{Z}_n^{ijl}\textrm{Z} _n^{\dagger \; mok}}{\omega-(E_0^N-E_n^{N-1})-i\eta},
\end{align}
in which
\begin{align}\label{G3basis_e:eq}
    \textrm{X}_n^{ijl}\!=\!\langle \Psi_0^N|\hat c^{\dagger}_l\hat c_j\hat c_i|\Psi_n^{N+1}\rangle & &\textrm{X}_n^{\dagger \; mok}\!=\!\langle \Psi_n^{N+1}|c^{\dagger}_m\hat c^{\dagger}_o \hat c_k \hat  |\Psi_0^N\rangle \nonumber \\
    \textrm{Z}_n^{ijl} = \langle\Psi_n^{N-1} |\hat c^{\dagger}_l\hat c_j\hat c_i|\Psi_0^N\rangle && \textrm{Z} _n^{\dagger \; mok}=\langle \Psi_0^N|c^{\dagger}_m\hat c^{\dagger}_o\hat c_k\hat |\Psi_n^{N-1}\rangle.
\end{align}
From the spectral representation of the 3-GF in Eq.~\eqref{G3basis:eq} we see that the poles of the 3-GF are equal to those of the 1-GF.
Instead, the corresponding amplitudes, given in Eq.~\eqref{G3basis_e:eq}, are different.
\subsection{The non-interacting 3-GF}
To understand the space in which Eq.~\eqref{Dyson:eq} has to be solved we first analyse the non-interacting 3-GF. 
Using the Wick theorem~\cite{wick50} and the transformations in Eqs.~\eqref{projection:eq}, it becomes
%
\begin{align}\label{noninteractingG3:eq}
    G^{0}_{ijl;mok}(\omega)&= G^0_{i;o}(\omega)G^0_{j;l}G^0_{m;k} - G^0_{j;o}(\omega)G^0_{i;l}G^0_{m;k} \nonumber \\
    &+G^0_{j;m}(\omega)G^0_{i;l}G^0_{o;k}-G^0_{i;m}(\omega)G^0_{j;l}G^0_{o;k}\nonumber \\
    &+[G^0_{i;m}G^0_{j;o}G^0_{l;k}](\omega)-[G^0_{i;o}G^0_{j;m}G^0_{l;k}](\omega),
\end{align}
where $[G^0_1G^0_1G^0_1](\omega)$ implies a frequency convolution, and $G^0_1=G^0_1(\tau=0^-)$.
The spectral representations of the contributions to Eq.~\eqref{noninteractingG3:eq} are given by
\begin{align}
   G^0_{i;o}(\omega)G^0_{j;l}G^0_{m;k}  &= - \frac{\delta_{io}\delta_{jl}\delta_{mk}f_jf_m}{\omega-\epsilon_i^0+\text{sign}(\epsilon^0_i-\mu)i \eta }\label{G0basis:eq}\\
    [G^0_{i;m}G^0_{j;o}G^0_{l;k}](\omega)&=
    \frac{\delta_{im}\delta_{jo}\delta_{lk}(f_i-f_l)(f_j-f_l)}{\omega-\epsilon^0_i-(\epsilon^0_j-\epsilon^0_l)+i\eta\text{sign}(\epsilon^0_i-\mu)},
    \label{convG0:eq}
\end{align}
in which $\epsilon^0_i$ and $f_i$ are the energy and occupation number corresponding to $\phi_i$, and where we used that $G^0_{ij}=i\delta_{ij}f_i$.
We note that the occupation numbers $(f_i-f_l) (f_j-f_l)$ in Eq.~\eqref{convG0:eq} restrict $G_{3}^0$ to its $2e1h$ and $2h1e$ contributions~\cite{Riv22}. The indices $i,j,m,o$ refer to conduction (valence) states and they describe the $2e$ ($2h$) process, while $l$ and $k$ refer to valence (conduction) states and they describe the $h$ ($e$) process. 
While the poles of Eq.~\eqref{G0basis:eq} contain approximate quasi-particle energies, the poles of Eq.~\eqref{convG0:eq} contain approximate satellite energies.
We will therefore refer to these terms as quasiparticle and satellite contributions, respectively.
These contributions do not couple in $G^0_3$, i.e., in its representation in Eq.~\eqref{noninteractingG3:eq} $G^0_3$ is block diagonal with a quasi-particle and a satellite block.
It can be verified that Eq.~\eqref{noninteractingG3:eq} satisfies the following symmetry relations,
\begin{align}
    \label{G3symmetry:eq}
    G^0_{ijl;mok}&=-G^0_{jil;mok}=-G^0_{ijl;omk}=G^0_{jil;omk}
    \\
    G^0_{ivv;mv'v'}(\omega) &= G^0_{i;m}(\omega) \quad (i\neq v, m \neq v' \quad \forall\,v,v')
\end{align}
where $v$ and $v'$ refer to occupied orbitals, i.e., $f_v = f_{v'} = 1$.
\footnote{We note that the symmetry in Eq.~\eqref{G3symmetry:eq} is also fulfilled by the interacting $G_3$ as can be seen from Eq.~\eqref{G3basis:eq}.}
As a consequence $G^0_3(\omega)$, as defined in Eq.~\eqref{noninteractingG3:eq}, is singular.
It contains redundant information that can be removed without loss of generality.

After removing all the redundant information, the non-interacting 3-GF can be written in the following matrix representation
\begin{equation}\label{G0matrix:eq}
    G^0_{3}(\omega)=\begin{pmatrix}
        G^0_{1}(\omega) & 0 \\
        0 & G_3^{0,3\text p}(\omega)
    \end{pmatrix},
\end{equation}
%
where
\begin{align}
    G^0_{i;m}(\omega)&=\frac{\delta_{im}}{\omega-\epsilon^0_i+i\eta\text{sign}(\epsilon^0_i-\mu)}, \\ 
    G_{i>jl;m>ok}^{0,3 \text p}(\omega)&=
    \frac{\delta_{im}\delta_{jo}\delta_{lk}(f_i-f_l)(f_j-f_l)}{\omega-\epsilon^0_i-(\epsilon^0_j-\epsilon^0_l)+i\eta\text{sign}(\epsilon^0_i-\mu)}.
    \label{G033p:eq}
\end{align}
Equation~\eqref{G0matrix:eq} thus defines the space in which the multichannel Dyson equation has to be solved.
For describing direct and inverse photoemission, the main idea is to use the three-body channel $G_3^{0,3 \text p}(\omega)$ as a reservoir that is coupled with the 1-body channel through a multichannel self-energy.

All the relations we have obtained so far still hold if the non-interacting 3-GF is replaced with an independent-particle 3-GF. In particular, it is convenient to replace $G^0_{3}(\omega)$ with the Hartree-Fock 3-GF. In this way, Hartree and exchange contributions to the self-energy are already included in Eq.~\eqref{G0matrix:eq}, and the multichannel self-energy has to take into account only the correlation part. 
Therefore, in the following, it will be understood that $G^0_{3}(\omega)$ refers to the Hartree-Fock 3-GF.
\subsection{The multichannel self-energy}
With the partition of $G^0_{3}$ given in Eq.~\eqref{G0matrix:eq}, the Dyson equation in Eq.~\eqref{Dyson:eq} becomes a multichannel Dyson equation, in which the multichannel self-energy is defined as
\begin{align}\label{selfmatrix:eq}
    \Sigma_3=\begin{pmatrix}
        0 & \Sigma^{\text c}\\
        \tilde \Sigma^{\text c} & \Sigma^{3\text p}
    \end{pmatrix}.
\end{align}
The MCDE can be represented diagrammatically as
\begin{widetext}
\begin{equation}\label{Eqn:Dyson_diag}
    \begin{gathered}
            \includegraphics[width=0.98\textwidth,clip=]{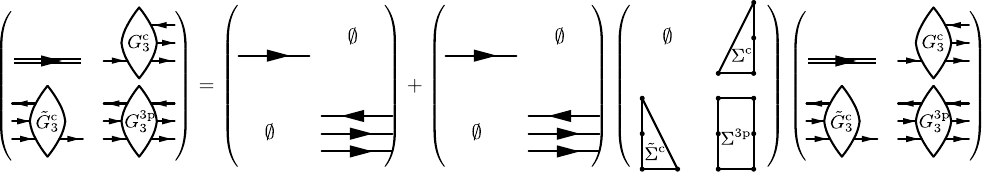} 
    \end{gathered}   
\end{equation}
\end{widetext}

From Eq.~(\ref{Eqn:Dyson_diag}), it is evident that the 3-GF consists of the 1-GF (\raisebox{-0.3\totalheight}{\includegraphics[width=0.1\textwidth]{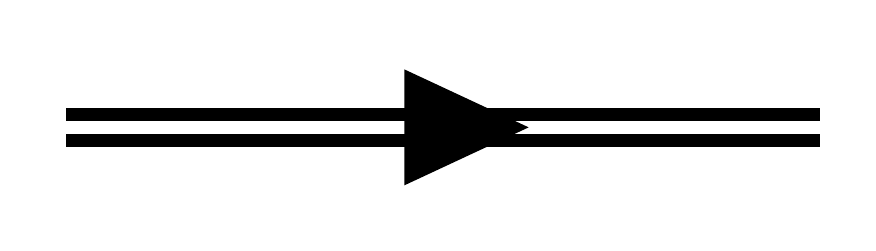}}) along with an explicit three-body component $G_3^{3\text p}$ (\raisebox{-0.3\totalheight}{\includegraphics[width=0.03\textwidth]{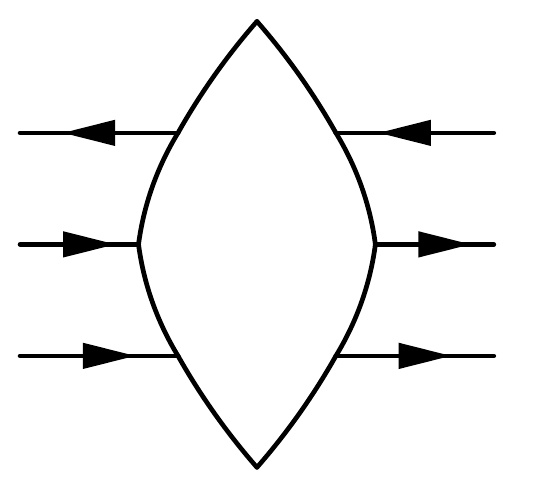}}), and the coupling between the 1-GF and $G_3^{3\text p}$, denoted as $G^c_3$ (\raisebox{-0.3\totalheight}{\includegraphics[width=0.03\textwidth]{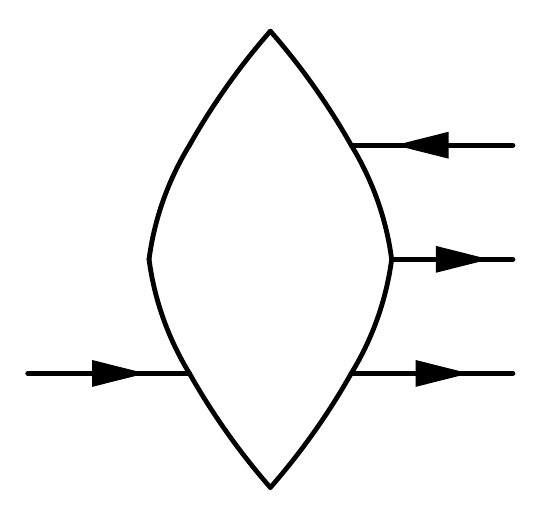}}) and $\tilde{G}^c_3$ (\raisebox{-0.3\totalheight}{\includegraphics[width=0.03\textwidth]{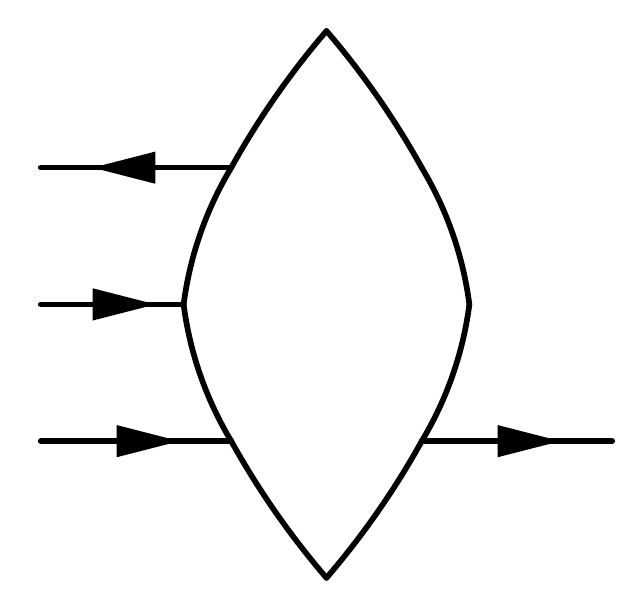}}). We represent the self-energy coupling terms, $\Sigma^{c}_{ijl,m}$ and $\tilde{\Sigma}^c_{i,mok}$, and the body of the self-energy, $\Sigma^{3\text p}_{ijl,mok}$,  using right triangles and a rectangle, respectively, to reflect their dimensions.

For practical calculations, an approximation to $\Sigma_3$ is required. 
To achieve this, we only consider correlations between pairs of particles in $\Sigma^{3 \text p}$ with each pair both having a direct and an exchange interaction. We thus treat each interaction between particles at the RPA+exchange (RPAx) level.
The coupling terms $\Sigma^{\text c}$ and $\tilde{\Sigma}^{\text c}$ are four-point quantities and, therefore, correspond to two-particle channels.
For consistency, we also treat these interactions at the RPAx level. 
As we will show, since second- and higher-order contributions in the interaction are introduced by $\Sigma^{\text c}$, $\tilde{\Sigma}^{\text c}$ and $\Sigma^{3 \text p}$, the head of $\Sigma_3$ vanishes since $G^0_3$ is already exact at first order.
Since our approximation to the multichannel self-energy is based on the RPAx approximation for the two-particle correlator, it is instructive to analyze the kernel of the Bethe-Salpeter equation for the 2-GF. 
We note that, alternatively, approximations for $\Sigma_3$ might be obtained by enforcing exact constraints such as conservation laws \cite{Baym_1961}.
\section{The two-body Green's function}
\label{sec:MCSE}

At zero temperature, the time-ordered equilibrium 2-GF is defined as
\begin{equation}\label{G2def:eq}
    G(1,2,1',2')=-\langle\Psi_0^N|T[\hat \psi(1)\hat \psi(2)\hat \psi^{\dagger}(2')\hat\psi^{\dagger}(1')]|\Psi_0^N\rangle.
\end{equation}
Making use of Wick's theorem, it is possible to show that the 2-GF can be decomposed as~\cite{Fetter}
\begin{align}\label{G2parts:eq}
        G(1,2,1',2')&=G(1,1')G(2,2')-G(1,2')G(2,1')\nonumber \\
        &+\delta G(1,2,1',2'),
\end{align}
where the first two terms on the right-hand side describe the independent propagation of two dressed particles, while $\delta G_2$ describes the interaction between these particles. In the context of the 2-GF, we refer to the $G_1G_1$ terms as non-interacting terms. Here, ``interaction" specifically denotes the interaction between the two particles.

The 2-GF given in Eq.~\eqref{G2def:eq} depends on four distinct times, which can be reduced to three time differences (if the Hamiltonian is time independent). In the majority of cases, it is used to study the simultaneous propagation of two particles (electrons or holes) or one electron and one hole. As a result, only the time difference corresponding to this propagation is relevant. The other two time differences account for the creation and annihilation of the electron-hole pair or the particle-particle pair. In most cases, these processes are considered to be instantaneous and, therefore, these time differences are set to zero. Let us now analyse separately the electron-hole and particle-particle channel.

\subsection{\textit{eh} channel}\label{sec:ph_channel}

To describe the electron-hole contribution, we set $t_1=t_{1'}^+$ and $t_2=t_{2'}^+$ in Eq.~\eqref{G2def:eq}. 
The $eh$ 2-GF hence becomes
\begin{align}\label{G2ehdef:eq}
    G^{\text{eh}}&(1,2,1',2')=\nonumber \\
    &-\langle\Psi_0^N|\hat{T}[(\hat \psi(x_1)\hat \psi^{\dagger}(x_{1'}))_{t_1}(\hat \psi(x_2)\hat\psi^{\dagger}(x_{2'}))_{t_2}]|\Psi_0^N\rangle,
\end{align}
where the notation $(\hat \psi(x_1)\hat \psi^{\dagger}(x_{1'}))_{t_1}$ implies that the two field operators both act at the time $t_1$.
\footnote{It is important to note that the other choice of ordering the field operators, namely $(\hat \psi(x_1)\hat \psi^{\dagger}(x_{2'}))_{t_1}(\hat \psi(x_2)\hat\psi^{\dagger}(x_{1'}))_{t_2}$, is also valid.} This choice of the times describes the propagation of an electron-hole pair. The calculation of the electron-hole channel is well-established and it can be found in literature (see, e.g., Refs.~[\onlinecite{csanak1971, Stri88}]). We will present here only the final result. In frequency space, it reads
\begin{align}\label{spectrlG2eh:eq}
    G^{\text{eh}}&(x_1,x_2,x_{1'},x_{2'};\omega)=-i\! \lim_{\eta \to 0^+}\! \sum_{n=0}^{\infty} \nonumber \\
    &\!\left[\frac{\chi_n(x_1,x_{1'})\tilde{\chi}_n(x_2,x_{2'})}{\omega-(E_n^N-E_0^N)+i\eta}\! - \! \frac{\tilde{\chi}_n(x_1,x_{1'})\chi_n(x_2,x_{2'})}{\omega+(E_n^N-E_0^N)-i\eta}\right] 
\end{align}
where the electron-hole amplitudes are defined as
\begin{align}
    \chi_n(x_1,x_{1'})&=\langle\Psi_0^N|\hat \psi(x_1)\hat \psi^{\dagger}(x_{1'})|\Psi_n^N\rangle, \\
    \tilde\chi_n(x_1,x_{1'})&=\langle\Psi_n^N|\hat \psi(x_1)\hat \psi^{\dagger}(x_{1'})|\Psi_0^N\rangle.
\end{align}
The poles of the $eh$ channel of the 2-GF are the neutral excitation (and de-excitation) energies. 

Applying the choice of the times mentioned above to the two non-interacting terms, i.e., 
$G(1,1')G(2,2')$and $G(1,2')G(2,1')$,
we obtain
\begin{align}
    G(1,1')G(2,2')&=G(x_1,x_{1'};0^-)G(x_2,x_{2'};0^-), \label{GGstatic:eq} \\ G(1,2')G(2,1')&=G(x_1,x_{2'};t_1\!-\!t_2)G(x_2,x_{1'};t_2\!-\!t_1).\label{GGdinamic:eq}
\end{align}
%
We see that one non-interacting term is static while the other non-interacting contribution depends on $t_1 - t_2$.
It can be shown that the Fourier transform of Eq.~\eqref{GGdinamic:eq} has poles at the neutral excitation energies
of a non-interacting electron-hole pair.
To study the $eh$ channel of the 2-GF, the so-called two-particle correlation function $L$ is defined as
\begin{equation}\label{LG2:eq}
    L(1,2,1',2')=-G^{eh}(1,2,1',2')+G(1,1')G(2,2'),
\end{equation}
in which the static combination of $G_1G_1$ in Eq.\eqref{GGstatic:eq} is subtracted from $G_2^{\text{eh}}$. 
This corresponds to removing the term corresponding to $n=0$ in the sum of Eq.~\eqref{spectrlG2eh:eq}~\cite{csanak1971}. 

A closed equation for $L$ is given by the Bethe-Salpeter equation (BSE) which links the non-interacting 
two-particle function $L^0(1,2,1',2')= G(1,2')G(2,1')$ to $L$ according to
\begin{align}\label{BSE:eq}
    L&(1,2,1',2')=L^0(1,2,1',2')\nonumber \\
    &+L^0(1,3,1',3')\Xi(3',4',3,4)L(4,2,4',2')
\end{align}
where repeated variables are integrated, and $\Xi$ is the kernel of the BSE defined by
\begin{align}\label{Xidef:eq}
    \Xi(1',2',1,2)&\equiv \frac{\delta \Sigma(1',1)}{\delta G(2,2')}=\nonumber \\
    &=-i\delta(1,1')\delta(2,2')v(1,2)+ \frac{\delta \Sigma_{\text{xc}}(1',1)}{\delta G(2,2')},
\end{align}
in which $\Sigma=v_H+\Sigma_{xc}$ is the one-body self-energy, with $v_H$ and $\Sigma_{xc}$ the Hartree and exchange-correlation contributions, respectively.
The first term on the right-hand side results from the functional derivative of the Hartree potential with respect to $G_1$.

Diagrammatically the BSE can be represented as 
\begin{equation}
    \begin{gathered}
        \includegraphics[width=0.47\textwidth,clip=]{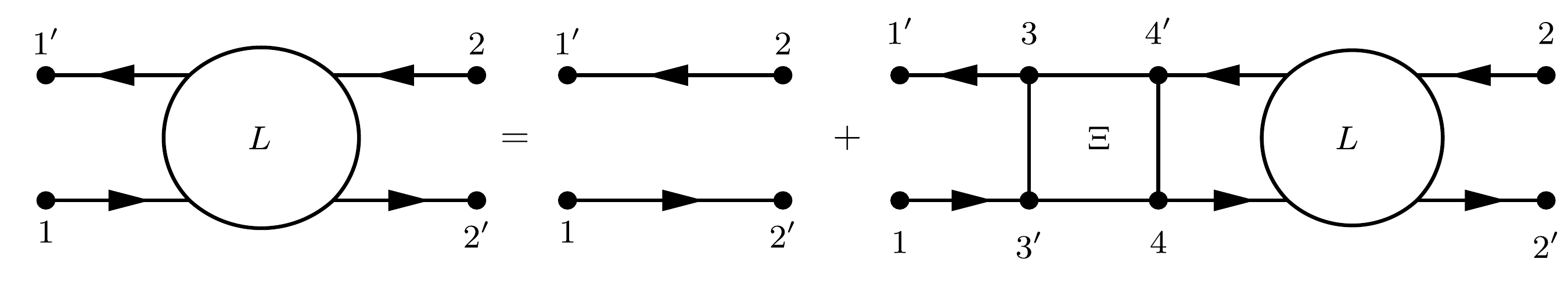}.
    \end{gathered}
\end{equation}
To obtain an expression suitable for practical calculations, we need to make approximations to $\Sigma_{\text{xc}}$. 
At first order in the interaction, it corresponds to the Fock exchange $\Sigma^F_{\text{xc}}(1',1)=iv(1',1)G(1',1)$. 
Inserting this approximation into Eq.~\eqref{Xidef:eq} we obtain the RPAx approximation to the kernel $\Xi$ according to
\begin{align}\label{XiHF:eq}
   \Xi^{\text{RPA}x}(1',2',1,2) =&-i\delta(1,1')\delta(2,2')v(1,2)\
   \nonumber \\ & +i\delta(1,2')\delta(2,1')v(1,2).
\end{align}
The first term on the right-hand side is the  $eh$ exchange interaction, while the second term is the  direct $eh$ interaction. 
Diagrammatically $\Xi^{\text{RPA}x}$ can be represented as
\begin{equation}\label{ppBSE-RPAx}
    \begin{gathered}
        \includegraphics[width=0.47\textwidth,clip=]{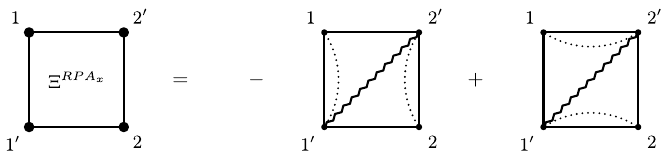},
    \end{gathered}
\end{equation}
where a dotted line represent a Dirac delta function and a wiggly line represents the bare Coulomb interaction.

%
It is instructive to analyze the $eh$ exchange and direct $eh$ contributions separately.
Let us first write the BSE with the $eh$ exchange kernel only and iterate it. We obtain
%
%
\begin{align}\label{screeningfromBSE:eq}
    &L^{\text{RPA}}(1,2,1',2') = L^0(1,2,1',2')-iL^0(1,3,1',3)\nonumber \\
    &\times v(3,4)L^0(4,2,4,2')-L^0(1,3,1',3)v(3,4)\nonumber\\
    &\times L^0(4,5,4,5)v(5,6)L^0(6,2,6,2')+...= \nonumber\\
    &L^0(1,2,1',2')  -iL^0(1,3,1',3)W^{\text{RPA}}(3,4)L^0(4,2,4,2').
\end{align}
where $W^{\text{RPA}}$ is the screened Coulomb potential in the random-phase approximation (RPA).
It is given by
\begin{equation}
W^{\text{RPA}}(1,2) = v(1,2) -i v(1,3) L^0(3,4,3,4) W^{\text{RPA}}(4,2)
\end{equation}
Its diagrammatic representation is given by
\begin{equation}\label{LH:fig}
    \begin{gathered}
        \includegraphics[width=0.47\textwidth,clip=]{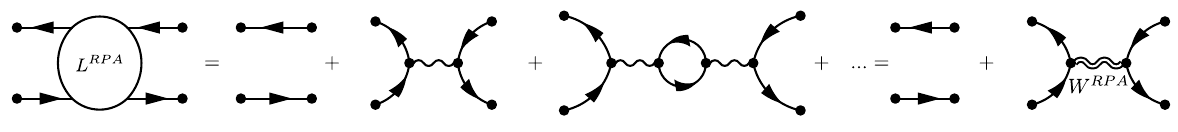}.
    \end{gathered}
\end{equation}

The analysis above clearly shows that the $eh$ exchange interaction is naturally screened in the BSE within the RPAx approximation.

\begin{figure}
\centerline{
\includegraphics[width=0.4\textwidth,clip=]{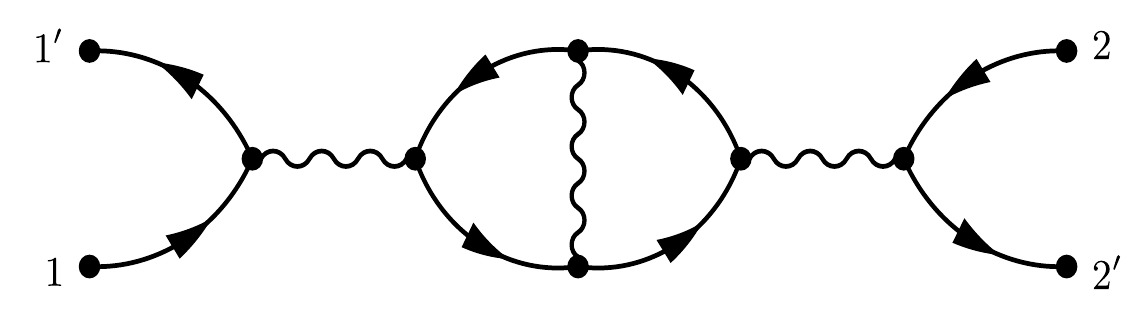}\hfill
}
\caption{\label{third:fig} A third-order term of the BSE with the $\Xi^{\text{RPA}x}$ or the $\Xi^{GW}$ kernel. This diagram shows a term of the screening beyond the RPA.}
\end{figure}

Let us now iterate the BSE with the kernel only containing the direct $eh$  interaction.
The diagrammatic representation of this iteration is given by
%
\begin{equation}
    \begin{gathered}
        \includegraphics[width=0.47\textwidth,clip=]{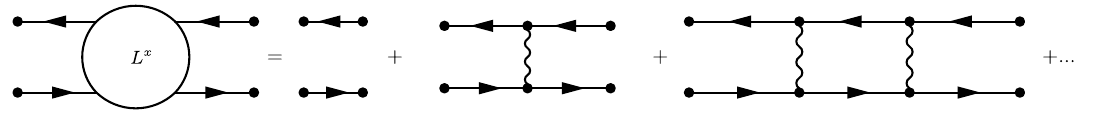}.
    \end{gathered}
\end{equation}
We see that in this case the interaction is not naturally resummed and the direct $eh$ interaction is unscreened in the RPAx approximation.
Screening of the direct $eh$ interaction can be achieved by replacing the bare Coulomb interaction by the screened Coulomb interaction in the second term on the right-hand side of Eq.~\eqref{XiHF:eq}.
We thus obtain

%
%
\begin{align}\label{XiGW:eq}
   &\Xi^{GW}(1',2',1,2) =\nonumber \\ &
   -i\delta(1,1')\delta(2,2')v(1,2)+i\delta(1,2')\delta(2,1')W(1,2).
\end{align}
This expression can also be obtained by using the $GW$ approximation to the self-energy in Eq.~\eqref{Xidef:eq} and 
neglecting the functional derivative $\delta W/ \delta G_1$, as is usually done, since this term is of second order in $W$.
In practice, $W^{\text{RPA}}$ is often used for the screened interaction in Eq.~\eqref{XiGW:eq}.
We note that both $\Xi^{\text{RPA}x}$ and $\Xi^{GW}$ introduce vertex corrections to the RPA screening but only for the $eh$ exchange term, e.g., the diagram reported in Fig.~\ref{third:fig}.
This shows that for the $eh$ exchange interaction screening beyond the RPA is naturally included. 

For completeness we also report the BSE with the RPAx kernel in orbital space. 
By using Eq.\eqref{projection:eq} we obtain
\begin{align}\label{Ltransform:eq}
 L_{ij;mo}(\omega)&=\int d x_1d x_{1'}dx_2 dx_{2'} \phi^*_{i}(x_1)\phi_{j}(x_{1'}) \nonumber \\ &\times  L(x_1,x_2,x_{1'},x_{2'};\omega)\phi^*_{m}(x_2)\phi_o(x_{2'}).
\end{align}
Therefore, the BSE in orbital space is
\begin{align}
    L_{ij;mo}(\omega)=L^0_{ij;mo}(\omega)+L^0_{ij;i'j'}(\omega)\Xi^{\text{RPA}x}_{i'j';m'o'}L_{m'o';mo}(\omega)
\end{align}
with the RPAx kernel equal to
\begin{equation}
    \Xi^{\text{RPA}x}_{ij;mo}=-iv_{imoj}+iv_{imjo}
\end{equation}
in which
\begin{equation}\label{potential:eq}
    v_{ijom}=\int dx_1 dx_2 \phi^*_i(x_1)\phi^*_j(x_2)v(\mathbf{r}_1,\mathbf{r}_2)\phi_o(x_2)\phi_m(x_1).
\end{equation}
Let us now turn our attention to the $pp$ channel.


%

\subsection{\textit{pp} channel}

To describe the propagation of two particles, i.e., two electrons or two holes, we choose the times in Eq.~\eqref{G2def:eq} as $t_1=t_2^+$ and $t_{1'}=t_{2'}^+$.
The particle-particle 2-GF is thus given by 
\begin{align}\label{G2ppdef:eq}
    G^{\text{pp}}&(1,2,1',2')= \nonumber \\
    &\langle\Psi_0^N|T[(\hat \psi(x_1)\hat \psi(x_2))_{t_1}(\hat \psi^{\dagger}(x_{1'})\hat\psi^{\dagger}(x_{2'}))_{t_{1'}}]|\Psi_0^N\rangle,
\end{align}
and its spectral representation is
\begin{align}\label{spectralG2pp:eq}
    &G^{\text{pp}}(x_1,x_2,x_{1'},x_{2'};\omega)= i\lim_{\eta \to 0^+}\sum_n \nonumber \\
    &\left[\frac{\zeta_n(x_1,x_2)\zeta^*_n(x_{2'},x_{1'})}{\omega-(E_n^{N+2}-E_0^N)+i\eta}- \frac{\tilde{\zeta}_n(x_{1'},x_{2'})\tilde\zeta^*_n(x_2,x_1)}{\omega+(E_n^{N-2}-E_0^N)-i\eta}\right] 
\end{align}
where the $pp$ amplitudes are defined as
\begin{align}
    \zeta_n(x_1,x_2)&=\langle\Psi_0^N|\hat \psi(x_1)\hat \psi(x_2)|\Psi_n^{N+2}\rangle, \\
    \tilde\zeta_n(x_{1'},x_{2'})&=\langle\Psi_0^N|\hat \psi^{\dagger}(x_{1'})\hat \psi^{\dagger}(x_{2'})|\Psi_n^{N-2}\rangle, 
\end{align}
and $\zeta^*_n$ and $\tilde\zeta^*_n$ are the corresponding complex conjugates.
The poles of Eq.~\eqref{spectralG2pp:eq} are the excitation energies corresponding to the removal or addition of two electrons.
For the choice of times mentioned above the non-interacting terms in Eq.~\eqref{G2parts:eq} become
\begin{align}
    G(1,1')G(2,2')&=G(x_1,x_{1'};t_1-t_{1'})G(x_2,x_{2'};t_1-t_{1'}) \label{GGppdirect:eq} \\ G(1,2')G(2,1')&=G(x_1,x_{2'};t_1-t_{1'})G(x_2,x_{1'};t_1-t_{1'}).\label{GGppinverse:eq}
\end{align}
We see that for the $pp$ channel, both non-interacting terms depend on $t_1-t_{1'}$. 
It can be shown that the Fourier transforms of both combinations have poles at energies corresponding to the addition of two non-interacting electrons and to the addition of two non-interacting holes.
Therefore, the non-interacting term for this channel is composed of both combinations, i.e.,
\begin{equation}\label{G0pp}
 G^{0\text{pp}}(1,2,1',2') = G^0(1,1')G^0(2,2') - G^0(1,2')G^0(2,1').
\end{equation}
In analogy with the $eh$ BSE this would suggest the following $pp$ BSE equation
\begin{align}\label{ppBSE:eq}
    &G^{\text{pp}}(1,2,1',2')=G^{0\text{pp}}(1,2,1',2')\nonumber \\
    &+G^{0\text{pp}}(1,2,3',4')K(3',4',3,4)G^{\text{pp}}(3,4,1',2'),
\end{align}
which defines $K$ as the kernel of the particle-particle channel.
It is useful to represent the $pp$ BSE~\eqref{ppBSE:eq} in its diagrammatic form 
\begin{equation}
    \begin{gathered}
            \includegraphics[width=0.47\textwidth,clip=]{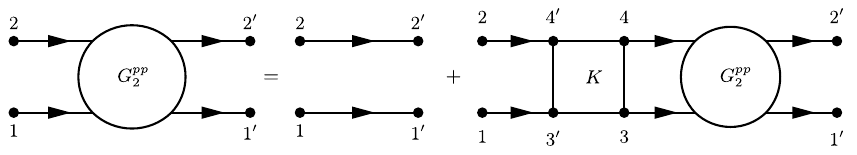} 
    \end{gathered},
    \label{G2pp:fig}
\end{equation}
In the (static) RPAx approximation this kernel is written as
\begin{align}\label{kernelpputilize:eq}
     K^{\text{RPA}x}(1',2',1,2)&=i\delta(1,1')\delta(2,2')v(1,2)
\nonumber \\&-i\delta(1,2')\delta(2,1')v(1,2).
\end{align}
Diagrammatically, this approximation can be represented as
\begin{equation}\label{pp-RPAX-kernel}
    \begin{gathered}
        \includegraphics[width=0.47\textwidth,clip=]{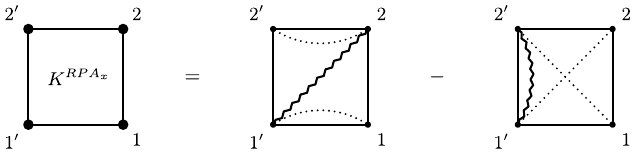}.
    \end{gathered}
\end{equation}
This expression can be compared to the RPAx kernel of the $eh$ BSE given in Eq.~\eqref{ppBSE-RPAx}.
Here, the analysis of both terms is similar to that of the direct $eh$ interaction of the $eh$ BSE, i.e., iterating the $pp$ BSE in Eq.~\eqref{G2pp:fig} will not screen the interactions.
Screening of the interaction in both terms can be achieved by replacing the bare Coulomb interaction by the screened Coulomb interaction.

It is instructive to project Eq.~\eqref{G0pp} on the basis that diagonalizes $G^0_1$.
By using the change of basis in Eq.~\eqref{projection:eq}, the matrix elements of the $pp$ 2-GF are
\begin{align}
    G^{\text{pp}}_{ij;mo}(\omega)&=\int dx_1 dx_{1'} dx_2 dx_{2'} \phi_i^*(x_1)\phi_j^*(x_2) \nonumber \\
    & \times G^{\text{pp}}(x_1,x_2,x_{1'},x_{2'};\omega)\phi_m(x_{1'})\phi_o(x_{2'}).
\end{align}
The non-interacting part thus becomes
\begin{align}\label{G0ppnon:eq}
      G^{0\text{pp}}_{ij;mo}(\omega) &= [G^0_{i;m}G^0_{j;o}](\omega)-[G^0_{i;o}G^0_{j;m}](\omega)\nonumber \\ &
      =-i \frac{(\delta_{im}\delta_{jo}-\delta_{io}\delta_{jm})(1-f_i-f_j)}{\omega-(\epsilon^0_i+\epsilon^0_j)+i\eta \text{sign}(\epsilon^0_i-\mu)}.
\end{align}
From the above equation one can verify that the following symmetry relations hold
\begin{equation}\label{G2symmetry:eq}
    G^{0\text{pp}}_{ij;mo}=-G^{0\text{pp}}_{ji;mo}=-G^{0\text{pp}}_{ij;om}=G^{0\text{pp}}_{ji;om}.
\end{equation}
This means that $G^{0\text{pp}}(\omega)$, as defined in Eq.~\eqref{G0ppnon:eq}, is singular.
As a consequence Eq.~\eqref{ppBSE:eq} cannot be solved by inversion.

However, as was the case for $G_3^0$, the redundant information can be removed without loss of generality.
We thus redefine $G^{0\text{pp}}$ as
\begin{align}\label{G0ppnonnonsing:eq}
      G^{0\text{pp}}_{i>j;m>o}(\omega) &
      =-i \frac{\delta_{im}\delta_{jo}(1-f_i-f_j)}{\omega-(\epsilon^0_i+\epsilon^0_j)+i\eta \text{sign}(\epsilon^0_i-\mu)}.
\end{align}
We can now write the $pp$ BSE as
\begin{align}
    G^{\text{pp}}_{ij;mo}(\omega) & = G^{0\text{pp}}_{ij;mo}(\omega)
    \nonumber \\&+
    G^{0\text{pp}}_{ij;i'j'}(\omega) K^{\text{RPA}x}_{i'j';m'o'}G^{\text{pp}}_{m'o';mo}(\omega),
\end{align}
where, for notational convenience, the restrictions $i>j$ and $m>o$ have been made implicit, and
%
%
%
\begin{equation}\label{kernelppvtilde:eq}
    K^{\text{RPA}x}_{ij;mo}=i v_{ijom}-i v_{ijmo}.
\end{equation}
\subsection{The multichannel self-energy}
As mentioned before, to obtain an approximation for the body of the multichannel self-energy we let the particles interact pairwise within the RPAx approximation, i.e., the self-energy contains all first-order direct and exchange interactions.
With 2 electrons and 1 hole (or 2 holes and 1 electron) there are 5 RPAx contributions, each containing one direct and one exchange interaction.
One contribution accounts for the particle-particle interaction while four contributions are needed to account for all electron-hole interactions, since there are two electron-hole pairs and for each of these pairs either the electron or the hole can exchange with the third particle.
The coupling terms are four-point kernels and are also approximated at the RPAx level.
It can be verified that they are equivalent to the RPAx kernel of the $pp$ BSE given in Eq.\eqref{pp-RPAX-kernel}.
We thus obtain the following approximation to the multichannel self-energy,
\begin{align}
    \Sigma^{3 \text p}_{ijl;mok}&=\!\! [(1\!-\!f_i)\!(1\!-\!f_j)f_l\!-\!f_if_j(1\!-\!f_l)][\delta_{lk} \bar v_{ijom} \nonumber \\ &+\!\delta_{mj}\bar v_{iklo} \!+\! \delta_{io} \bar v_{jklm} \!-\! \delta_{oj}  \bar v_{iklm} \!-\! \delta_{im}  \bar v_{jklo}] ,\label{selfthird:eq}\\
    \Sigma^{\text c}_{i;mok}&=\bar v_{ikom}, \label{selfcoupling:eq} \\
     \tilde \Sigma^{\text c}_{ijl;m}&=\bar v_{ijlm}, \label{selfcouplingtilde:eq}
\end{align}
where $\bar v_{ikom}=v_{ikom}-v_{ikmo}$ with $v_{ikom}$ defined in Eq.~\eqref{potential:eq}.
%
The occupation numbers in Eq.~\eqref{selfthird:eq} guarantee that $\Sigma^{3 \text p}$ has opposite signs for the $2e1h$ channel and the $2h1e$ channel.
It is important to note that $\Sigma^{3\text p}$ is static and hermitian.
Moreover, the $2h1e$ and $2e1h$ channel are uncoupled.
Therefore, $\Sigma^{3\text p}$ is block-diagonal with two blocks $\Sigma^{2e1h}$ and $\Sigma^{2h1e}$.
We note that there are other approaches that obtain a similar self-energy, in particular the 2-particle-hole Tamm-Dancoff approximation~\cite{Sch78} (see also Refs.~[\onlinecite{Dei16,Tor19}] in the context of trions).
\section{Diagrammatic analysis of the MCDE}
\label{sec:analysis}
To analyze the diagrammatic structure of the MCDE in Eq.~\eqref{Eqn:Dyson_diag} containing the multichannel self-energy given in Eqs.~\eqref{selfthird:eq},~\eqref{selfcoupling:eq} and~\eqref{selfcouplingtilde:eq} it is convenient to represent it in real space. 
Applying the change of basis given in Eq.~\eqref{G3changebasis:eq} to the multichannel self-energy yields the following expression.
\begin{align}
     \Sigma^{3 \text p}_{ijl;mok}&\!=\! \int\! dx_1 dx_2 dx_3 dx_{1'} dx_{2'} dx_{3'} \phi^*_i(x_1)\phi^*_j(x_2)\phi_l(x_{3'})  \nonumber\\
     &\!\!\!\!\!\!\!\! \times  \Sigma^{3\text p}(x_1 , x_2 , x_{3'} ,x_{1'} , x_{2'} , x_3)\phi_m(x_{1'})\phi_o(x_{2'})\phi^*_k(x_3), 
      \label{selfchange:eq}\\ 
     \Sigma^{\text c}_{i;mok} &=  \int dx_1  dx_3 dx_{1'} dx_{2'}   \Sigma^{\text c}(x_1 ,x_{1'} , x_{2'} , x_3) \nonumber \\
      &\phi^*_i(x_1) \phi_m(x_{1'})\phi_o(x_{2'})\phi^*_k(x_3), \label{selfchangec:eq}
     \\ 
     \tilde \Sigma^{\text c}_{ijl;m} &= \int dx_1 dx_2  dx_{1'} dx_{3'}  \tilde \Sigma^{\text c}(x_1 , x_2 , x_{3'} ,x_{1'} ) \nonumber \\
      &\phi^*_i(x_1)\phi^*_j(x_2)\phi_l(x_{3'}) \phi_m(x_{1'}). \label{selfchangect:eq}
\end{align}
For the approximation given in Eq.~\eqref{selfthird:eq} we thus obtain
%

\begin{align}
    &\Sigma^{2e1h}\!(x_1 ,\! x_2 ,\! x_{3'} ,\! x_{1'} ,\! x_{2'} ,\! x_3)\!=\!-\Sigma^{2h1e}(x_1 ,\! x_2 ,\! x_{3'} ,\! x_{1'} ,\! x_{2'} ,\! x_3)\nonumber \\
    &=\!\delta(x_3,\!x_{3'})[\delta(x_1,\!x_{1'}\!)\delta(x_2,\!x_{2'}\!)\!-\!\delta(x_1,\!x_{2'}\!)\delta(x_2,\!x_{1'}\!)]v(\mathbf{r}_1,\!\mathbf{r}_2)\nonumber\\
    &-\!\delta(x_{1'},\!x_2)[\delta(x_1,\!x_{3'}\!)\delta(x_3,\!x_{2'}\!)\!-\!\delta(x_1,\!x_{2'}\!)\delta(x_3,\!x_{3'}\!)]v(\mathbf{r}_1,\!\mathbf{r}_3)\nonumber  \\
    &-\!\delta(x_1,\!x_{2'})[\delta(x_2,\!x_{3'}\!)\delta(x_3,\!x_{1'}\!)\!-\!\delta(x_2,\!x_{1'}\!)\delta(x_3,\!x_{3'}\!)]v(\mathbf{r}_2,\!\mathbf{r}_3)\nonumber \\
    &+\!\delta(x_2,\!x_{2'})[\delta(x_1,\!x_{3'}\!)\delta(x_3,\!x_{1'}\!)\!-\!\delta(x_1,\!x_{1'}\!)\delta(x_3,\!x_{3'}\!)]v(\mathbf{r}_1,\!\mathbf{r}_3)\nonumber \\
    &+\!\delta(x_1,\!x_{1'})[\delta(x_2,\!x_{3'}\!)\delta(x_3,\!x_{2'}\!)\!-\!\delta(x_2,\!x_{2'}\!)\delta(x_3,\!x_{3'}\!)]v(\mathbf{r}_2,\!\mathbf{r}_3)\label{selfbodyreal:eq}
\end{align}
and 
\begin{align}
    &\Sigma^{\text c}(x_1 ,x_{1'} , x_{2'} , x_3)=\nonumber \\&=[\delta(x_1,x_{1'})\delta(x_{2'},x_3)-\delta(x_1,x_{2'})\delta(x_{1'},x_3)]v(\mathbf{r}_1,\mathbf{r}_3) \label{selfcouplingreal:eq} \\ 
    &\tilde \Sigma^{\text c}(x_1 , x_2 , x_{3'} ,x_{1'} )=\nonumber \\&=[\delta(x_1,x_{1'})\delta(x_2,x_{3'})-\delta(x_1,x_{3'})\delta(x_2,x_{1'})]v(\mathbf{r}_1,\mathbf{r}_2).\label{selfcouplingtildereal:eq}
\end{align}

From these expressions in real space, it becomes easier to understand the diagrammatic structure. 
The $2e1h$ contribution to the body in Eq.~\eqref{selfbodyreal:eq} is given by 
\begin{widetext}  
\begin{equation}
    \begin{gathered}
            \includegraphics[width=0.98\textwidth,clip=]{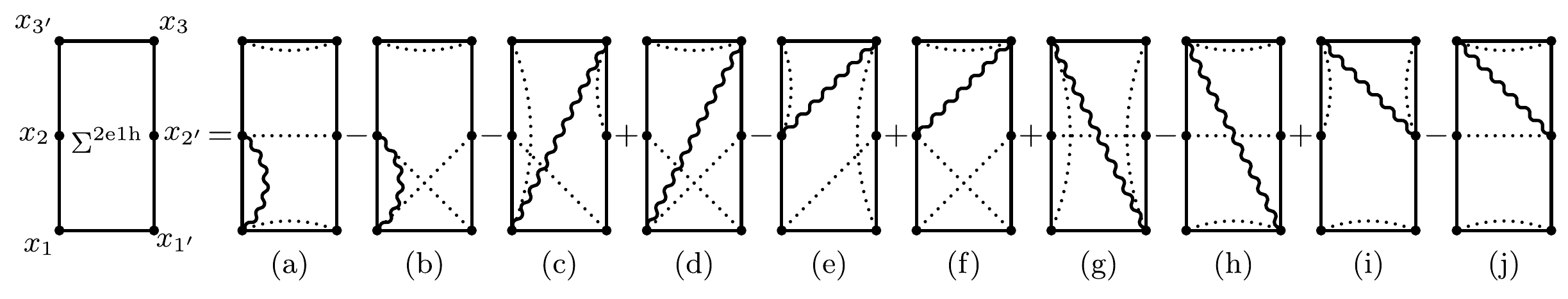} 
    \end{gathered}.
    \label{self_body:fig}
\end{equation}
\end{widetext}
The $2h1e$ diagrams are equal to the $2e1h$ diagrams in Eq.~\eqref{self_body:fig} except for an overall minus sign.
The coupling terms are given by
\begin{align}
    \begin{gathered}
        \includegraphics[width=0.48\textwidth,clip=]{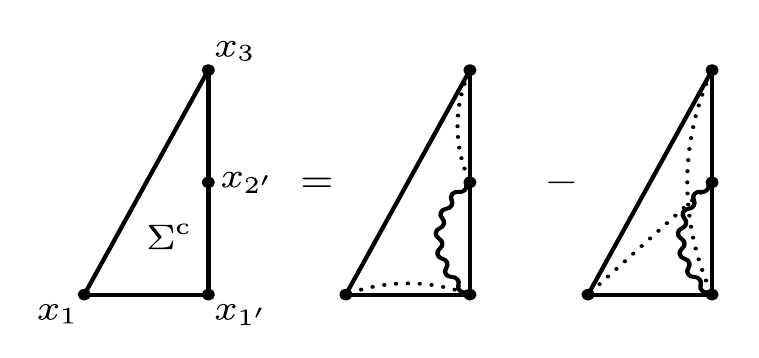}
    \end{gathered}
        \label{selfcoupling:fig} \\  
    \begin{gathered}
        \includegraphics[width=0.47\textwidth,clip=]{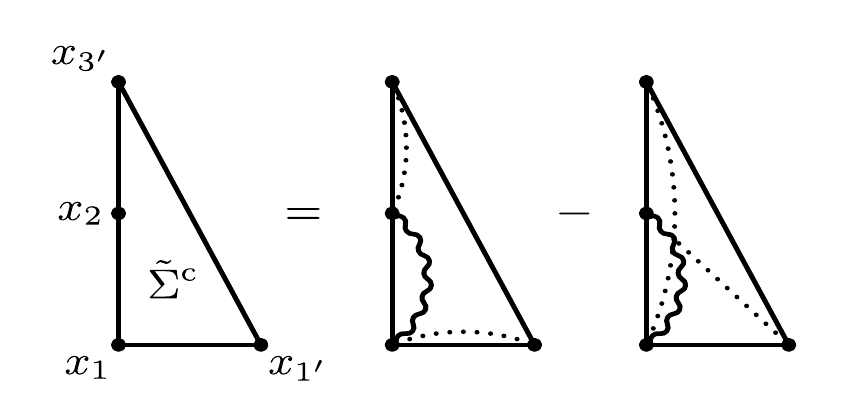}
    \end{gathered}      
    \label{selfcouplingtilde:fig}              
\end{align}
which correspond to Eq.~\eqref{selfcouplingreal:eq}, and Eq.~\eqref{selfcouplingtildereal:eq}, respectively.

\begin{figure}
\centerline{
\includegraphics[width=0.2\textwidth,clip=]{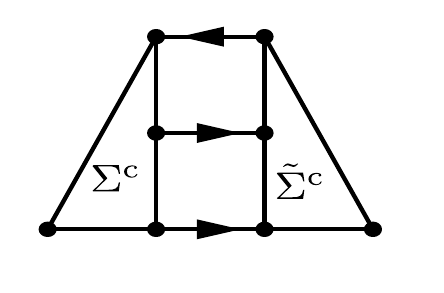}\hfill
\includegraphics[width=0.24\textwidth,clip=]{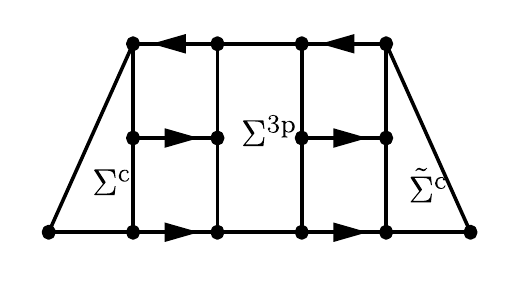}\hfill
}
\caption{\label{selfsecondthird:fig} Diagrammatic representation of a general second- and third-order 1-body self-energy obtained by iterating the MCDE~\eqref{Eqn:Dyson_diag}. By inserting multichannel self-energy diagrams, Eqs.~\eqref{self_body:fig}-\eqref{selfcouplingtilde:fig}, all the second- and third-order proper skeleton 1-body self-energy diagrams are obtained.}
\end{figure}

\begin{figure}
\centerline{
\includegraphics[width=0.5\textwidth,clip=]{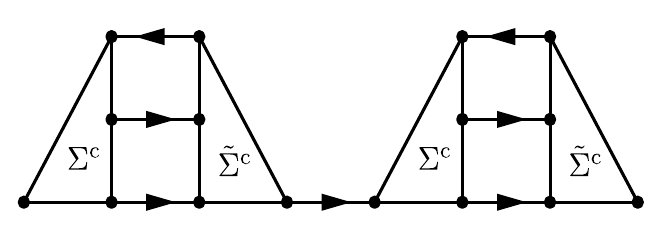}
}
\caption{\label{multi_improper:fig} An improper fourth-order diagram created by the repetition of two second-order diagrams.}
\end{figure}

To understand which diagrams are added to $G_1^0$ when solving the MCDE, one can iterate Eq.~\eqref{Eqn:Dyson_diag} and inspect the head of the final matrix which corresponds to $G_1(\omega)$. The first iteration does not change the head since we include the contributions that are of first order in the interaction already in $G^0_{3}(\omega)$. In Fig.~\ref{selfsecondthird:fig} we show the general structure of the one-body self-energy obtained from the MCDE after the second (left-hand side of Fig.~\ref{selfsecondthird:fig}) and third iterations (right-hand side of Fig.~\ref{selfsecondthird:fig}). 
In general, $n$ iterations of the MCDE yield one-body self-energy proper skeleton diagrams of $n$th order in the interaction.
To construct these diagrams, one has to insert $n-1$ $G_{3}^{0,3\text p}$ elements and $n-2$ $\Sigma^{3\text p}$ rectangles between $\Sigma^c$ and $\tilde\Sigma^c$ with the $G_{3}^{0,3\text p}$ and $\Sigma^{3\text p}$ insertions alternating.
We note that iterating the MCDE in Eq.~\eqref{Eqn:Dyson_diag} also creates improper diagrams such as the fourth-order diagram depicted in Fig.~\ref{multi_improper:fig}. 


%


%
%


As mentioned before $G^0_{3}(\omega)$ refers to the Hartree-Fock 3-GF.
Therefore, each $G^0_1$ line contained in the $G_3^{0,\text 3p}$ diagram already has HF self-energy insertions.
One can include higher-order diagrams by dressing $G_3^{0,\text 3p}$ beyond HF using, e.g., second Born, $GW$ or the T-matrix within the quasiparticle approximation. In this case, it is useful to assume that the correlated $G^{0,3 \text p}_{3}$ is diagonal in the same basis as $G_3^0$. This approximation is also used, for example, to derive the cumulant approximation \cite{Guz11,martin_reining_ceperley_2016}.

With this approach, we have already shown~\cite{Riv23} that we are exact at second order and have provided examples of how to obtain third and fourth order diagrams. 
Here we would like to analyse
how screening effects are naturally included in the MCDE.

To understand this point, we focus our attention on the following diagram of the one-body self-energy that is second order in the interaction, namely
\begin{align}
    \begin{gathered}
                   \includegraphics[width=0.42\textwidth,clip=]{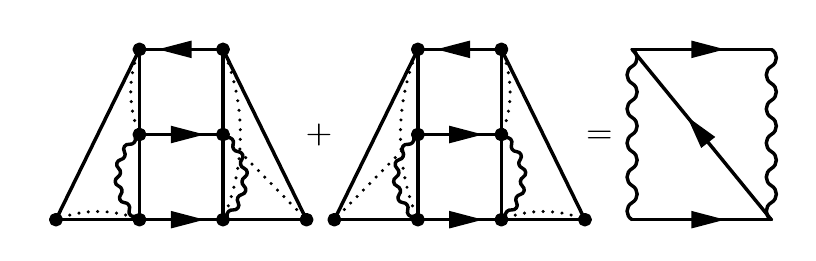}  
    \end{gathered}
\end{align}
Because of the conditions $i>j$ and $m>o$ applied to $G_3^{0,3\text p}(\omega)$ in Eq.~\eqref{G033p:eq} the second order diagram on the right-hand side is the sum of the two MCDE diagrams of the left without there being any double counting.~\cite{Riv23}
More details are given in section \ref{sec:diagrams}.
%

We will now show how this diagram can be resummed when combined with higher-order diagrams obtained from the MCDE.
For example, the MCDE gives rise to the following third-order diagram that has a polarization bubble,
\begin{equation}
    \begin{gathered}
            \includegraphics[width=0.47\textwidth,clip=]{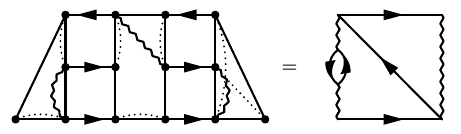} 
    \end{gathered},
    \label{screening_left:fig}
\end{equation}
%
which can be obtained from the general diagram on the right-hand side of Fig.~\ref{selfsecondthird:fig} 
with the multichannel self-energy component (i) given in Eq.~\eqref{self_body:fig} 
and the first and second terms on the right-hand side of Eqs.~\eqref{selfcoupling:fig} and ~\eqref{selfcouplingtilde:fig}, respectively.
We note that in reality the diagram on the right-hand side of Eq.~\eqref{screening_left:fig} is the sum of 4 MCDE diagrams.
For notational convenience we reported only one of those diagrams.
In a similar way we can obtain diagrams with 2 bubbles, 3 bubbles etc.
All those diagrams can be resummed into the following diagram
\begin{equation}
    \begin{gathered}
            \includegraphics[width=0.1\textwidth,clip=]{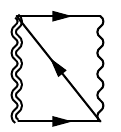} 
    \end{gathered}.
\end{equation}

In a similar way we can resum other diagrams to obtain 
\begin{equation}
    \begin{gathered}
            \includegraphics[width=0.1\textwidth,clip=]{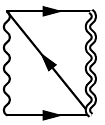} 
    \end{gathered}.
\end{equation}

We note that these resummations are similar to that of the $eh$ exchange interaction in the $eh$ BSE performed in Eqs.~\eqref{screeningfromBSE:eq} and~\eqref{LH:fig}.


Finally, we want to highlight that in our method screening beyond the RPA is included.
For example, using, diagrams (i) and (j) of the multichannel self-energy in Eq.~\eqref{self_body:fig}, we obtain 
\begin{equation}
    \begin{gathered}
            \includegraphics[width=0.47\textwidth,clip=]{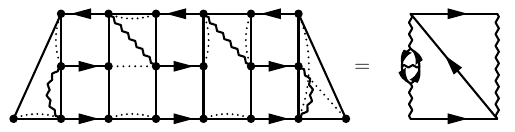} 
    \end{gathered}
    \label{screening_left_plusRPA:fig}
\end{equation}
where, once more, the MCDE diagram on the left represents a sum of several diagrams.

In general, the MCDE naturally includes screening beyond the RPA for every interaction in the coupling terms ~\eqref{selfcoupling:eq} and~\eqref{selfcouplingtilde:eq} and in the $eh$ exchange interactions in $\Sigma^{3\text p}$, i.e., diagrams (c), (e), (g) and (i) of Eq.~\eqref{self_body:fig}. 
The interactions in the other diagrams of Eq.~\eqref{self_body:fig} remain unscreened.
For example, in Eq.~\eqref{screening_left_plusRPA:fig} the interaction inside the bubble will remain unscreened because it comes from the diagram (j) in Eq.~\eqref{self_body:fig}. 
This means that we can further improve $\Sigma_3$ by dressing the Coulomb interactions appearing in the diagrams (a), (b), (d), (f), (h) and (j) of $\Sigma_3^{3p}$ in Eq.~\eqref{self_body:fig}, for example, by using a screened Coulomb interaction.
%
\section{Evaluating diagrams of the MCDE}\label{calculate:sec}
\label{sec:diagrams}
In this section we show how to formally compare diagrams obtained from the MCDE in Eq.~\eqref{Eqn:Dyson_diag} to those obtained from perturbation theory. For simplicity, we report here only the diagrams, whereas all the details of the derivation are given in  Appendix~\ref{AppendxA}.
\subsection{Second-order diagrams}
We start from the second-order direct and exchange diagrams obtained from perturbation theory, which are depicted on the right-hand side of Eqs.~\eqref{direct:fig} and ~\eqref{vertexxy:fig}, respectively.  
These second-order contributions are both found in the MCDE one-body self-energy diagram on the left-hand side of Fig.~\ref{selfsecondthird:fig}. 
Indeed, inserting the diagrams corresponding to $\Sigma_c$ and $\tilde{\Sigma}_c$, given in Eqs ~\eqref{selfcoupling:fig} and ~\eqref{selfcouplingtilde:fig}, respectively, we obtain four diagrams.
Two diagrams combine only exchange or only direct contributions and together they yield the direct second-order diagram as
\begin{equation}\label{direct:fig}
    \begin{gathered}
        \includegraphics[width=0.47\textwidth,clip=]{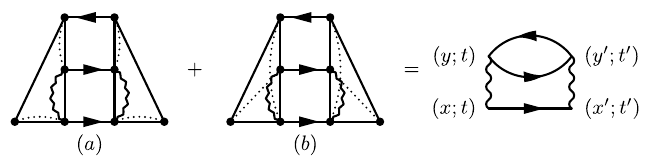}.
    \end{gathered}
\end{equation}

The remaining two diagrams, which combine an exchange and a direct contribution, together yield the exchange second-order diagram as
\begin{equation}\label{vertexxy:fig}
    \begin{gathered}
        \includegraphics[width=0.48\textwidth,clip=]{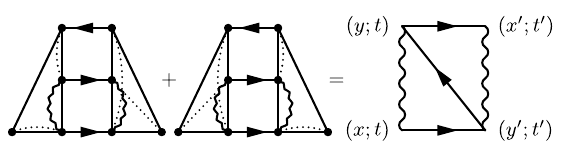}
    \end{gathered}.
\end{equation}
We note that the space, spin, and time indices reported on the right-hand side of Eqs (\ref{direct:fig})-(\ref{vertexxy:fig}) refer to Eqs (\ref{2O_D})-(\ref{2O_X}) in Appendix~\ref{AppendxA}.
The fact that each second-order self-energy diagram is produced by two MCDE diagrams is due to the orbital space in which  $\Sigma_c$ and $\tilde{\Sigma}_c$ are defined. This can be better understood from the corresponding equations, as we show in Appendix~\ref{AppendxA}.
\subsection{Third-order diagrams}
To illustrate how third-order one-body self-energy terms are obtained from the MCDE we focus, as an example, on the ladder contribution obtained from perturbation theory, which is depicted on the right-hand side of Eq.~\eqref{third1:fig}. 
 \begin{equation}\label{third1:fig}
     \begin{gathered}
         \includegraphics[width=0.48\textwidth,clip=]{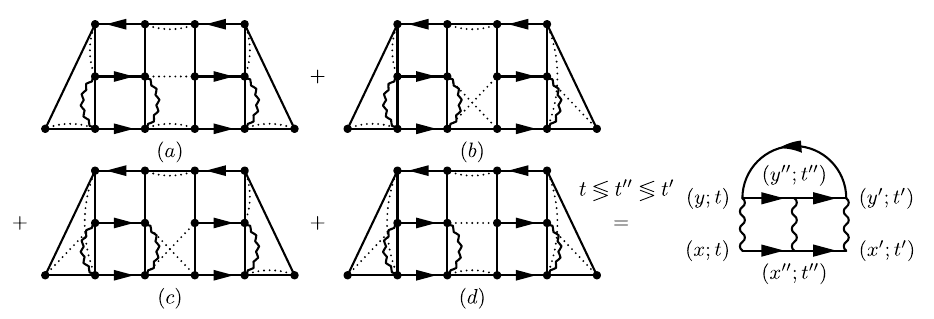},
     \end{gathered}
 \end{equation}
where the space, spin, and time indices on the right-hand side refer to Eq.~\eqref{Diag:ladder}.
This third-order contribution is also found in the MCDE one-body self-energy diagram on the right-hand side of Fig.~\ref{selfsecondthird:fig} with the constraint that $t < t''< t'$ or $t > t''> t'$.
This restriction on the times is indicated in Eq.~\eqref{third1:fig} by the expression $t\lessgtr t''\lessgtr t'$ above the equal sign, i.e., the equation holds if and only if $t < t''< t'$ or $t > t''> t'$.
This constraint results from our approximation to $\Sigma^{3 \text p}$ which is based on the kernels of the $pp$ and $eh$ channels of the BSE for the 2-GF. As a consequence, the $eeh$ and $hhe$ contributions to $\Sigma^{3 \text p}$ are not coupled. This seems a reasonable approximation since these two contributions describe different physical processes. The contribution (a), (b), (c), and (d) reported on the left-hand side of Eq.~\eqref{third1:fig} have similar expressions, but with different constraints on the orbitals. Therefore the sum of the four contributions cover the full orbital space. More details are given in Appendix~\ref{AppendxA}.

Another example of third-order contribution to the one-body self-energy is the vertex diagram represented by
\begin{equation}\label{third2:fig}
    \begin{gathered}
        \includegraphics[width=0.48\textwidth,clip=]{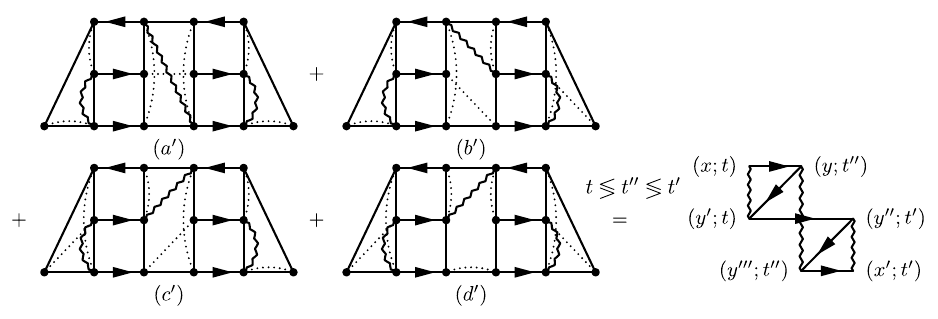},
    \end{gathered}
\end{equation}
where the space, spin, and time indices on the right-hand side refer to Eq.~\eqref{Diag:vertex}.
As for the ladder diagram, the MCDE yields the same diagram but with the time constraint $t\lessgtr t''\lessgtr t'$.
Similar time restrictions are also obtained for higher orders.
%
\section{Effective three particle Hamiltonian}
\label{sec:Hamiltonian}
In the previous sections we have given details about the derivation of the MCDE as well as an analysis of the diagrams that are included when using an RPAx approximation for the MCDE self-energy.
In this section we will show how the MCDE can be solved using standard numerical techniques.
Since the MCDE self-energy is static and Hermitian it is convenient to rewrite the MCDE in terms of an eigenvalue problem with an effective Hamiltonian. We can invert the MCDE according to $G_3^{-1} = [G_{3}^0]^{-1} - \Sigma_3$ and, therefore, we can write
\begin{widetext}
\begin{align}
[G_3(\omega)]^{-1} &= \begin{pmatrix}
       (\omega - \epsilon^0_i) \delta_{im} & \Sigma^c_{i;mok} \\
        \tilde\Sigma^c_{ijl;m} & \frac{(\omega - \epsilon^0_i-(\epsilon^0_j-\epsilon^0_l)) \delta_{im}\delta_{jo}\delta_{lk} - (f_{i}-f_{l}) (f_{j}-f_{l})\Sigma_{ijl;mok}^{3p}}{(f_{i}-f_{l}) (f_{j}-f_{l})}
    \end{pmatrix}
    \\ &=
    \begin{pmatrix}
       (\omega - \epsilon^0_i) \delta_{im} & (f_{m}-f_{k}) (f_{o}-f_{k})\Sigma^c_{i;mok} \\
        \tilde\Sigma^c_{ijl;m} & (\omega - \epsilon^0_i-(\epsilon^0_j-\epsilon^0_l)) \delta_{im}\delta_{jo}\delta_{lk} - (f_{i}-f_{l}) (f_{j}-f_{l})\Sigma_{ijl;mok}^{3p}
    \end{pmatrix}
    \begin{pmatrix}
       \delta_{im} & 0\\
        0& \frac{\delta_{im}\delta_{jo}\delta_{lk}}{(f_{i}-f_{l}) (f_{j}-f_{l})}
    \end{pmatrix}.
\end{align}
\end{widetext}
Using $[AB]^{-1} = B^{-1}A^{-1}$ we obtain

\begin{widetext}
\begin{align}
    G_3(\omega) &=
        \begin{pmatrix}
       \delta_{im} & 0\\
        0& (f_{i}-f_{l}) (f_{j}-f_{l})\delta_{im}\delta_{jo}\delta_{lk}
    \end{pmatrix}
    \nonumber \\&\times
    \begin{pmatrix}
       (\omega - \epsilon^0_i) \delta_{im} & (f_{m}-f_{k}) (f_{o}-f_{k})\Sigma^c_{i;mok} \\
        \tilde\Sigma^c_{ijl;m} & (\omega - \epsilon^0_i-(\epsilon^0_j-\epsilon^0_l)) \delta_{im}\delta_{jo}\delta_{lk} - (f_{i}-f_{l}) (f_{j}-f_{l})\Sigma_{ijl;mok}^{3p}
    \end{pmatrix}^{-1}
\end{align}
\end{widetext}
We can now define the following effective three-particle Hamiltonian
\begin{equation}
    H_3^{\text{eff}}=\begin{pmatrix}
        H^{1 \text p} & H^c  \\
        \tilde{H}^c  & H^{3 \text p}
    \end{pmatrix},
\end{equation}
where
\begin{align}
    H^{1\text p}_{i;m}&=  \epsilon^0_i \delta_{im}
    \\
    H^c_{i;m>ok}&= (f_{m}-f_{k}) (f_{o}-f_{k}) \Sigma^c_{i;mok},
    \\
    \tilde{H}^c_{i>jl;m}& = \tilde\Sigma^c_{ijl;m},
    \\
    H^{3\text p}_{i>jl;m>ok}&= (\epsilon^0_{i}-(\epsilon^0_{l}-\epsilon^0_{j})) \delta_{im}\delta_{jo}\delta_{lk}\nonumber\\
    &+ (f_{i}-f_{l}) (f_{j}-f_{l}) \Sigma_{ijl;mok}^{3p}.
\end{align}

We can thus rewrite $G_3$ as
\begin{equation}
G_3(\omega) \! = \! \begin{pmatrix}
       \delta_{im} & 0\\
        0& (f_{i}-f_{l}) (f_{j}-f_{l})\delta_{im}\delta_{jo}\delta_{lk}
    \end{pmatrix} \!\! \left[\mathbf{1}\omega \! -\! H^{\text{eff}} \right]^{-1} .  
\end{equation}
The occupation numbers $(f_i-f_l) (f_j-f_l)$ restrict $G_{3}$ to its $2e1h$ and $2h1e$ contributions.
It is therefore convenient to define a modified effective Hamiltonian according to
\begin{equation}\label{Heff:eq}
    \bar{H}_3^{\text{eff}}=\begin{pmatrix}
        H^{1\text p} & H^c  \\
        \bar{\tilde{H}}^c  & H^{3\text p}
    \end{pmatrix},
\end{equation}
where
\begin{equation}
    \bar{\tilde{H}}^c_{i>jl;m} =  (f_{i}-f_{l})(f_{j}-f_{l}) \tilde\Sigma^c_{ijl;m}.
\end{equation}
%
%

We can now write
\begin{equation}
\bar{H}_3^{\text{eff}} A_{\lambda} = E_{\lambda}A_{\lambda},
\end{equation}
where $E_{\lambda}$ and $A_{\lambda}$ are the eigenvectors and eigenvalues of $\bar{H}_3^{\text{eff}}$, respectively.
The spectral representation of $\left[\mathbf{1}\omega - \bar{H}_3^{\text{eff}} \right]^{-1}$ can thus be written as
%
\begin{equation}
\left[\mathbf{1}\omega - \tilde{H}_3^{\text{eff}} \right]^{-1}_{\mu\nu} = \sum_{\lambda} \frac{A^\mu_{\lambda}A^{*\nu}_{\lambda}}{\omega-E_{\lambda}}.
\end{equation}
We thus obtain the following expression for $G_3$,
%
\begin{equation}
G_3 (\omega) =
\begin{pmatrix}
        G_1(\omega) & G^c(\omega)  \\
        \tilde{G}^c(\omega)  & G^{3\text p}(\omega)
    \end{pmatrix},
\end{equation}
with
\begin{align}
    G_{i;m}(\omega)& =  \sum_{\lambda}\frac{A^i_{\lambda}A^{*m}_{\lambda}}{\omega-E_{\lambda}},
    \\
    G^c_{i;m>ok}(\omega)&=  \sum_{\lambda}\frac{A^i_{\lambda}A^{*mok}_{\lambda}}{\omega-E_{\lambda}},
    \\
    \tilde{G}^c_{i>jl;m}(\omega)& =  \sum_{\lambda}\frac{A^{ijl}_{\lambda}A^{*m}_{\lambda }}{\omega-E_{\lambda}},
    \\
    G^{3p}_{i>jl;m>ok}(\omega)&=  \sum_{\lambda}\frac{A^{ijl}_{\lambda}A^{*mok}_{\lambda}}{\omega-E_{\lambda}}.
\end{align}
The spectral function $A(\omega) = \frac{1}{\pi} |\textrm{Im} G(\omega)|$ is then given by
\begin{align}
A(\omega) &= \frac{1}{\pi} \sum_i \left| \textrm{Im}   G_{i;i}(\omega) \right| \nonumber\\
&= \frac{1}{\pi} \sum_i\left| \textrm{Im} \sum_{\lambda}\frac{A^i_{\lambda}A^{*i}_{\lambda }}{\omega-E_{\lambda}+i\eta \text{sgn}(E_{\lambda}-\mu)}\right|
\end{align}
where $\mu$ is the chemical potential.

The eigenvalue problem in Eq.~\eqref{Heff:eq} can be solved by direct diagonalisation or by more efficient iterative techniques.
Using the Haydock-Lanczos solver~\cite{Hay72,Schm03,Her05} we can directly solve for the spectral function.
The numerical scaling of the calculation will then be determined by the construction of $H^{3 \text p}$ 
which scales as  $N_v^4 N_c + N_c^4 N_v$, where $N_v$ and $N_c$ are the number of occupied and unoccupied states, respectively.
We note that in the case of solids the above scaling has to be multiplied with $N^4_{\mathbf{k}}$ where $N_{\mathbf{k}}$ is the number of $\mathbf{k}$-points.
The overall scaling with respect to the number of electrons $N$ is thus $N^5$.


\section{Conclusions}
\label{sec:conclusions}
We have given a detailed analysis of the multi-channel Dyson equation that we have recently proposed.~\cite{Riv23}.
In particular, we have explained the rationale behind the approximation to the multichannel self-energy 
by comparing it to the kernels of the electron-hole and particle-particle Bethe-Salpeter equations of the 2-body Green's function.
Furthermore, we have shown how one can extract diagrams from the multi-channel Dyson equation and how they compare to diagrams obtained from standard perturbation theory.
Our analysis showed the richness of the physics described by a simple static RPAx-like approximation to the multi-channel self-energy. We also discussed how it can be further improved without double counting diagrams. 

In this work we have concentrated on the multi-channel Dyson equation which couples the 1-body and 3-body Green's functions, but this is a general strategy which can be systematically extended to other couplings. For example one can couple the 2-body and 4-body Green's functions in order to describe double excitations for the electron-hole channel of the 2-body Green's function, or to extract an approximation beyond RPAx for the kernel of the particle-particle channel of the 2-body Green's function. We are currently working on these two extensions.
Our approach could also be generalized to treat inelastic scattering.~\cite{Cederbaum_2000,Cederbaum_2001,Ofir_2002}

Finally, we have given a formulation of the multi-channel Dyson equation in terms of an eigenvalue problem with an effective Hamiltonian which is useful for practical implementations.

\acknowledgments
We thank the French “Agence Nationale de la Recherche (ANR)” for financial support (Grant Agreements No. ANR-19-CE30-0011 and No. ANR-22-CE30-0027).
\appendix
\section{Diagrams}\label{AppendxA}

\begin{widetext}
In this appendix we give more details about the second and third-order one-body self-energy obtained from the MCDE and depicted in Eqs (\ref{direct:fig}-(\ref{third2:fig})
%
\subsection{Second-order direct diagram from the MCDE}
We start from the second-order direct ($\Sigma^{(2),D}_1$) and exchange ($\Sigma^{(2),X}_1$) diagrams obtained from perturbation theory, which read
\begin{align}\label{2O_D}
    &\Sigma^{(2),D}_1(xt,x't')=\nonumber \\
    &G^0(xt,x't')\! \int \! dydy'v(x,\!y)v(x'\!,\!y')G^0(yt,\!y't')G^0(y't'\!,\!yt)
\end{align}
and 
\begin{align}\label{2O_X}
    \Sigma^{(2),X}_1(xt,x't')&=-\int dydy' G^0(xt,y't')G^0(y't',yt)\nonumber \\ &\times G^0(yt,x't')v(x,y)v(y',x'),
\end{align}
and which are depicted on the right and side of Eqs~\eqref{direct:fig} and ~\eqref{vertexxy:fig}, respectively.  
The second-order one-body self-energy obtained from the MCDE reads
\begin{align}\label{selfproper2:eq}
    &\Sigma^{(2)}_{1,im}(\omega)=\!\!\!\!\!\!\!\!\!\!\!\! \sum_{\substack{m'o'k' \\i''j''l''\\m'>o',i''>j''}}\!\!\! \!\!\!\!\!\!\!\Sigma^c_{i;m'o'k'}[G^0_{m';i''}G^0_{o';j''}G^0_{k';l''}](\omega)\tilde\Sigma^c_{i''j''l'';m}.
\end{align}
Using ~\eqref{selfcoupling:eq} and~\eqref{selfcouplingtilde:eq} and performing the Fourier transform to the time domain gives
\begin{align}\label{selfproper_t:eq}
    \Sigma^{(2)}_{1,im}(\tau)&=\!\!\!\!\!\sum_{\substack{m'o'k' \\i''j''l''\\m'>o',i''>j''}}\!\!\!\!\!\!\!\bar{v}_{ik'o'm'}\nonumber\\
    &\times G^0_{m';i''}(\tau)G^0_{o';j''}(\tau)G^0_{k';l''}(-\tau)\bar{v}_{i''j''l''m}.
\end{align}
Let us focus on the contributions which give the direct self-energy term $\Sigma^{(2),D}_1$ in Eq.~\eqref{2O_D}. They read 
\begin{align}\label{selfproper_t:eq}
    \Sigma^{(2),(a+b)}_{1,im}(\tau)&=\!\!\!\!\!\sum_{\substack{m'o'k' \\i''j''l''\\m'>o',i''>j''}}\!\!\!\!\!\!\!
    G^0_{m';i''}(\tau)G^0_{o';j''}(\tau)G^0_{k';l''}(-\tau) \left[v_{ik'o'm'}v_{i''j''l''m}+v_{ik'm'o'}v_{i''j''ml''}\right],
\end{align}
where the superscripts $a$ and $b$ on the left-hand side refer to the to the contributions $(a)$ and $(b)$, respectively, on the left-hand side of Eq.~ (\ref{direct:fig}). Since we work in the basis which diagonalizes $G^0$, for $\tau>0$ we have $G^0_{m';i''}(\tau)G^0_{o';j''}(\tau)G^0_{k';l''}(-\tau)=G^0_{c;c}(\tau)G^0_{c';c'}(\tau)G^0_{v;v}(-\tau)$, and for $\tau<0$ $G^0_{m';i''}(\tau)G^0_{o';j''}(\tau)G^0_{k';l''}(-\tau)=G^0_{v;v}(\tau)G^0_{v';v'}(\tau)G^0_{c;c}(-\tau)$. 
We therefore arrive at
\begin{align}\label{selfproper_t_final:eq}
    \Sigma^{(2),(a+b)}_{1,im}(\tau)&=\sum_{\substack{cc'v\\c>c'}}
   G^0_{c;c}(\tau)G^0_{c';c'}(\tau)G^0_{v;v}(-\tau) \left[v_{ivc'c}v_{cc'vm}+v_{ivcc'}v_{cc'mv}\right]\nonumber\\&+\sum_{\substack{vv'c\\v>v'}}
  G^0_{v;v}(\tau)G^0_{v';v'}(\tau)G^0_{c;c}(-\tau) \left[v_{icv'v}v_{vv'cm}+v_{icvv'}v_{vv'mc}\right]\nonumber\\
    &=\sum_{\substack{cc'v\\c\neq c'}}
   G^0_{c;c}(\tau)G^0_{c';c'}(\tau)G^0_{v;v}(-\tau)v_{ivc'c}v_{cc'vm}+\sum_{\substack{vv'c\\v\neq v'}}
   G^0_{v;v}(\tau)G^0_{v';v'}(\tau)G^0_{c;c}(-\tau)v_{icv'v}v_{vv'cm},
\end{align}
where, in the last equality we used that
\begin{align}
&\sum_{\substack{cc'v\\c> c'}} 
   G^0_{c;c}(\tau)G^0_{c';c'}(\tau)G^0_{v;v}(-\tau)\left[v_{ivc'c}v_{cc'vm}+v_{ivcc'}v_{cc'mv}\right]=\sum_{\substack{cc'v\\c> c'}}
   G^0_{c;c}(\tau)G^0_{c';c'}(\tau)G^0_{v;v}(-\tau)v_{ivc'c}v_{cc'vm}\nonumber\\
   &+\sum_{\substack{cc'v\\c< c'}}
   G^0_{c;c}(\tau)G^0_{c';c'}(\tau)G^0_{v;v}(-\tau)v_{ivc'c}v_{c'cmv}=\sum_{\substack{cc'v\\c\neq c'}}
   G^0_{c;c}(\tau)G^0_{c';c'}(\tau)G^0_{v;v}(-\tau)v_{ivc'c}v_{cc'vm}.
 \end{align}   
Here, in the first equality, we interchanged $c$ and $c'$ in the second term, and in the last equality we used that $v_{cc'vm} = v_{c'cmv}$ . The same derivation can be done for the case $v>v'$. 

Projecting back to real space, Eq.~(\ref{selfproper_t_final:eq}) gives (\ref{2O_D}). A similar derivation can be done for the terms which give $\Sigma^{(2),X}_1$.
\subsection{Third-order ladder diagram from the MCDE}
We now focus on the ladder contribution
\begin{align}\label{Diag:ladder}
    &\Sigma^{(3),\text{ladder}}_1(x,x',t-t')\!=\!-(i)^3\!\int dy dy'dy''dx'' dt'' v(x,y) \nonumber \\
    &\times v(x'',y'') v(x',y')G^0(x,x'',t-t'') G^0(x'',x',t''-t')\nonumber \\
    &\times G^0(y,y'',t-t'')G^0(y'',y',t''-t')G^0(y',y,t'-t),
\end{align}
obtained from perturbation theory, and which is depicted on the right-hand side of (\ref{third1:fig}). 
This third-order contribution
is also found in the MCDE one-body self-energy diagram on the left-hand side of Fig.~ \ref{selfsecondthird:fig} under some time restriction, as we explain in the following. 

Our starting point is the third order one-body self-energy  obtained from the MCDE, which reads
\begin{align}\label{thirdgeneral:eq}
    \Sigma_{1(im)}^{(3)}(\omega)&=\!\!\!\!\!\!\!\!\! \sum_{\substack{m'o'k' i'j'l'\\m''o''k''i''j''l''\\m'>o',i'>j',m''>o'',i''>j''}} \!\!\!\!\!\!\!\!\!\! \Sigma^\text c_{i;m'o'k'}[G^0_{m';i'}G^0_{o';j'}G^0_{k';l'}](\omega)\nonumber \\   &\!\!\!\!\!\!\!\!\!\ \times
    \Sigma_{i'j'l';m''o''k''}^{\text{3p}}[G^0_{m'';i''}G^0_{o'';j''}G^0_{k'';l''}](\omega)\tilde \Sigma^\text c_{i''j''l'';m}.
\end{align}
Its Fourier transform back to time domain gives 
\begin{align}\label{thirdgeneral_tau:eq}
    \Sigma_{1(im)}^{(3)}(\tau)&=\!\!\!\!\!\!\!\!\! \sum_{\substack{m'o'k' i'j'l'\\m''o''k''i''j''l''\\m'>o',i'>j',m''>o'',i''>j''}} \!\!\!\!\!\!\!\!\!\!   
    \Sigma^\text c_{i;m'o'k'}\nonumber \\ &\times\int d\tau' G^0_{m';i'}(\tau')G^0_{o';j'}(\tau')G^0_{k';l'}(-\tau')\nonumber \\ &\times
    \Sigma_{i'j'l';m''o''k''}^{\text{3p}} G^0_{m'';i''}(\tau-\tau')G^0_{o'';j''}(\tau-\tau')\nonumber \\ &\times
    G^0_{k'';l''}(\tau'-\tau)\tilde \Sigma^\text c_{i''j''l'';m}.
\end{align}

Due to the structure of the coupling terms $\Sigma^c$ and $\tilde{\Sigma}^c$ and of the body part $\Sigma^{3 \text p}$, there are in total forty terms which contribute to the third-order one-body self-energy.

We analyse the term that is obtained from Eq.~\eqref{thirdgeneral_tau:eq} by setting
\begin{align}
 \Sigma^\text c_{i;m'o'k'}&=v_{ik'o'm'}, & 
 \tilde \Sigma^\text c_{i''j''l'';m}= v_{i''j''l''m},&\nonumber\\
  \Sigma_{i'j'l';m''o''k''}^{\text{3p}} &=[(1-f_{i'})(1-f_{j'})f_{l'}-f_{i'}f_{j'}(1-f_{l'})]\delta_{l'k''}  v_{i'j'o''m''} \nonumber.
\end{align}
We hence get
\begin{align}\label{MCDE:ladder-1}
   & \Sigma_{1(im)}^{(3),(a)}(\tau)=\!\!\!\!\!\!\!\!\! \sum_{\substack{m'o'k' i'j'l'\\m''o''k''i''j''l''\\m'>o',i'>j',m''>o'',i''>j''}} \!\!\!\!\!\!\!\!\!\!       
    v_{ik'o'm'}
   \int d\tau' G^0_{m'i'}(\tau')G^0_{o'j'}(\tau')G^0_{k'l'}(-\tau')\nonumber \\ &\times
  [(1-f_{i'})(1-f_{j'})f_{l'}-f_{i'}f_{j'}(1-f_{l'})]\delta_{l'k''} v_{i'j'o''m''}G^0_{m''i''}(\tau-\tau')G^0_{o''j''}(\tau-\tau')G^0_{k''l''}(\tau'-\tau)v_{i''j''l''m},
\end{align}
where the superscript $a$ on the left-hand side refers to the contribution $(a)$ on the left-hand side of Eq.~\eqref{third1:fig}.
We work in the basis which diagonalizes $G^0$. Due to the occupation numbers in $\Sigma^{3 \text p}$ we can consider two cases: i) $i'=v, j'=v', l'=c$; ii)  $i'=c, j'=c, l=v$.

For case i) we get
\begin{align}\label{Eqn:vvc}
    \Sigma_{1(im)}^{(3),(a),-}(\tau)=  
   - \!\!\!\!\!\!\!\!\!\!\! \sum_{\substack{vv'v''v'''c\\v>v',v''>v'''}} \!\!\!\!\!\!\!\!\!\!  v_{icv'v} \!\int_\tau^0 \!\!\!\! d\tau' G^0_{v;v}(\tau')G^0_{v';v'}(\tau')G^0_{c;c}(-\tau')
 v_{vv'v'''v''}G^0_{v'';v''}(\tau-\tau')G^0_{v''';v'''}(\tau-\tau') G^0_{c;c}(\tau'-\tau)v_{v''v'''cm}.
\end{align}
Since $G^0_{v;v}(\tau')$, $G^0_{v';v'}(\tau')$, and $G^0_{c;c}(-\tau')$ are nonzero only for $\tau'<0$ and $G^0_{c;c}(\tau'-\tau)\neq 0$ for  $\tau'>\tau$, this implies that  $\tau<\tau'<0$. It moreover constrains $m''=v''$ and $o''=v'''$. The constraint over the time $\tau'$ can be reformulated as follows. Let us set $\tau=t-t'$. Since $\tau'$ is a generic time difference, we can set it as $\tau'=t''-t'$ and $d\tau'\rightarrow dt''$. Therefore we have the restriction $t<t''<t'$.

Similarly, for case ii) we get
\begin{align}\label{Eqn:ccv2}
    \Sigma_{1(im)}^{(3),(a),+}(\tau)=  \!\!\!\!\!\!\!\!\! \sum_{\substack{cc'c''c'''v\\c>c',c''>c'''}} \!\!\!\!\!\!\!\!\!\!   
    v_{ivc'c}\int^\tau_0 d\tau' G^0_{c;c}(\tau')G^0_{c';c'}(\tau')G^0_{v;v}(-\tau')
 v_{cc'c'''c''}G^0_{c'';c''}(\tau-\tau')G^0_{c''';c'''}(\tau-\tau') G^0_{v;v}(\tau'-\tau)v_{c''c'''vm},
\end{align}
The $G^0$ elements set the constraint $\tau'>0$ and $\tau'<\tau$, i.e. $0<\tau'<\tau$. This also constrains $m''=c''$ and $o''=c'''$. By setting $\tau=t-t'$ and $\tau'=t''-t'$ we get $t>t''>t'$. 

We now apply the relations \cite{Mar99}
$G^0_{c;c}(-\tau')G^0_{c;c}(\tau'-\tau)=-iG^0_{c;c}(-\tau)$ and $ G^0_{v;v}(-\tau')G^0_{v;v}(\tau'-\tau)=iG^0_{v;v}(-\tau)$ in (\ref{Eqn:vvc}) and (\ref{Eqn:ccv2}), respectively, which gives

\begin{align}\label{3a-}
    \Sigma_{1(im)}^{(3),(a),-}(\tau)= i  \!\!\!\!\!\!\!\!\! \sum_{\substack{vv'v''v'''c\\v>v',v''>v'''}} \!\!\!\!\!\!\!\!\!\!   
   v_{icv'v}\int d\tau' G^0_{v;v}(\tau')G^0_{v';v'}(\tau')
 v_{vv'v'''v''}G^0_{v''v''}(\tau-\tau')G^0_{v'''v'''}(\tau-\tau')
 G^0_{c;c}(-\tau)v_{v''v'''cm},
\end{align}
 and 
\begin{align}\label{3a+}
    \Sigma_{1(im)}^{(3),(a),+}(\tau)=    
   i\!\!\!\!\!\!\!\!\! \sum_{\substack{cc'c''c'''v\\c>c',c''>c'''}} \!\!\!\!\!\!\!\!\!\!v_{ivc'c}\int d\tau' G^0_{c;c}(\tau')G^0_{c';c'}(\tau')
 v_{cc'c'''c''}G^0_{c''c''}(\tau-\tau')G^0_{c'''c'''}(\tau-\tau')
 G^0_{v;v}(-\tau)v_{c''c'''vm}.
\end{align}

The other three terms $\Sigma_1^{(3),(b)}$, $\Sigma_1^{(3),(c)}$, and $\Sigma_1^{(3),(d)}$ are calculated with a similar expression as $\Sigma_1^{(3),(a)}$, where the Coulomb potential matrix elements have different indices. More specifically, the other terms are obtained from $\Sigma_1^{(3),(a)}$ with the following substitutions
\begin{align}
    \Sigma^{(3),(b)}_{1(im)}=\Sigma^{(3),(a)}_{1(im)} &  &\text{with} & & v_{i'j'o''m''}\to -v_{i'j'm''o''} &  &\text{and} & & v_{i''j''l''m}\to - v_{i''j''ml''} \nonumber \\
    \Sigma^{(3),(c)}_{1(im)}=\Sigma^{(3),(a)}_{1(im)} &  &\text{with} & & v_{i'j'o''m''}\to -v_{i'j'm''o''} &  &\text{and} & & v_{ik'o'm'}\to - v_{ik'm'o'}\nonumber \\
    \Sigma^{(3),(d)}_{1(im)}=\Sigma^{(3),(a)}_{1(im)} &  &\text{with} & & v_{ik'o'm'}\to - v_{ik'm'o'} &  &\text{and} & &  v_{i''j''l''m}\to - v_{i''j''ml''}. 
\end{align}
Using the same strategy as in Eqs(\ref{MCDE:ladder-1})-(\ref{3a+}) these differences create
\begin{align}
    \Sigma^{(3),(b)}_{1(im)}=\Sigma^{(3),(a)}_{1(im)} &  &\text{with} & & v''<v''' &  &\text{and} & & c''<c''' \nonumber \\
    \Sigma^{(3),(c)}_{1(im)}=\Sigma^{(3),(a)}_{1(im)} &  &\text{with} & & v'<v &  &\text{and} & & c'<c \nonumber \\
    \Sigma^{(3),(d)}_{1(im)}=\Sigma^{(3),(a)}_{1(im)} &  &\text{with} & &v''<v''',v'<v &  &\text{and} & &  c''<c''', c'<c
\end{align}
Therefore the sum of the four contributions cover the full orbital space. 

By summing $\Sigma^{(3),(a)}_1$, $\Sigma^{(3),(b)}_1$, $\Sigma^{(3),(c)}_1$, and $\Sigma^{(3),(d)}_1$ and projecting in real space we arrive at (\ref{Diag:ladder}) with the restriction that $t < t''< t'$ or $t > t''> t'$.
\subsection{Third-order vertex diagram from the MCDE}
Along the same lines as for the $\Sigma_1^{(3),\text{ladder}}$, we now show that the vertex contribution
\begin{align}\label{Diag:vertex}
    &\Sigma^{\text{(3),vertex}}(x,x',t-t')=(i)^3\int dydy'dy''dy'''dt'' v(x,y') \nonumber\\
    &\times v(y,y''')v(y'',x')G^0(x,y,t-t'')G^0(y,y'\!,t''-t) \nonumber\\
    &\times G^0(y',y''\!\!,t-t')G^0(y'',y'''\!,t'-t'')G^0(y'''\!,x',t''-t'),
\end{align}
depicted on the right-hand side of (\ref{third2:fig}) is recovered from the MCDE under some time constraints. Our starting point is Eq.~(\ref{thirdgeneral_tau:eq}). 
As an example we analyse the term ($a'$) reported on the left-hand side of (\ref{third2:fig}), which is obtained by setting
\begin{align}
 \Sigma^\text c_{i;m'o'k'}&=v_{ik'o'm'},\nonumber\\
 \tilde \Sigma^\text c_{i''j''l'';m}&= v_{i''j''l''m},\nonumber\\
  \Sigma_{i'j'l';m''o''k''}^{\text{3p}} &=[(1-f_{i'})(1-f_{j'})f_{l'}-f_{i'}f_{j'}(1-f_{l'})]\delta_{o''j'}  v_{i'k''m''l'} \nonumber.
\end{align}
We hence get
\begin{align}\label{MCDE:vertex-1}
    \Sigma_{1(im)}^{(3),(a')}(\tau)&=\!\!\!\!\!\!\!\!\! \sum_{\substack{m'o'k' i'j'l'\\m''o''k''i''j''l''\\m'>o',i'>j',m''>o'',i''>j''}} \!\!\!\!\!\!\!\!\!\!       
    v_{ik'o'm'}
    \int d\tau' G^0_{m'i'}(\tau')G^0_{o'j'}(\tau')G^0_{k'l'}(-\tau')\nonumber \\ &\times
  [(1-f_{i'})(1-f_{j'})f_{l'}-f_{i'}f_{j'}(1-f_{l'})]\delta_{o''j'}  v_{i'k''m''l'}G^0_{m''i''}(\tau-\tau')G^0_{o''j''}(\tau-\tau')G^0_{k''l''}(\tau'-\tau)v_{i''j''l''m}.
\end{align}
We work in the basis which diagonalizes $G^0$. Due to the occupation numbers in $\Sigma^{3 \text p}$ we can consider two cases: i) $i'=v, j'=v', l'=c$; ii)  $i'=c, j'=c', l=v$.

For case i) we get
\begin{align}\label{Eqn:vvc_vertex}
    \Sigma_{1(im)}^{(3),(a'),-}(\tau)=  
   - \!\!\!\!\!\!\!\!\! \sum_{\substack{vv'v''cc'\\v>v',v''>v'}} \!\!\!\!\!\!\!\!\!\!  v_{icv'v}\int_\tau^0 d\tau' G^0_{v;v}(\tau')G^0_{v';v'}(\tau')G^0_{c;c}(-\tau')
 v_{vc'v''c}G^0_{v''v''}(\tau-\tau')G^0_{v';v'}(\tau-\tau') G^0_{c';c'}(\tau'-\tau)v_{v''v'c'm}.
\end{align}
Since $G^0_{v;v}(\tau')$, $G^0_{v';v'}(\tau')$, and $G^0_{c;c}(-\tau')$ are nonzero only for $\tau'<0$ and $G^0_{c;c}(\tau'-\tau)\neq 0$ for  $\tau'>\tau$, this implies that  $\tau<\tau'<0$. It moreover constrains $m''=v''$ and $k''=c'$. The constraint over the time $\tau'$ can be reformulated as follows. Let us set $\tau=t-t'$. Since $\tau'$ is a generic time difference, we can set it as $\tau'=t''-t'$ and $d\tau'\rightarrow dt''$. Therefore we have the restriction $t<t''<t'$.

Similarly, for the case ii) we get
\begin{align}\label{Eqn:ccv_vertex}
    \Sigma_{1(im)}^{(3),(a'),+}(\tau)&=  \!\!\!\!\!\!\!\!\! \sum_{\substack{cc'c''vv'\\c>c',c''>c'}} \!\!\!\!\!\!\!\!\!\!  v_{ivc'c}\int_\tau^0 d\tau' G^0_{c;c}(\tau')G^0_{c';c'}(\tau')G^0_{v;v}(-\tau')
 v_{cv'c''v}G^0_{c''c''}(\tau-\tau')G^0_{c';c'}(\tau-\tau') G^0_{v';v'}(\tau'-\tau)v_{c''c'v'm}.
\end{align}
The $G^0$ elements set the constraint $\tau'>0$ and $\tau'<\tau$, i.e. $0<\tau'<\tau$. This also constrains $m''=c''$ and $k''=v'$. By setting $\tau=t-t'$ and $\tau'=t''-t'$ we get $t>t''>t'$. 

We now apply the relations \cite{Mar99}
$ G^0_{v';v'}(\tau')G^0_{v';v'}(\tau-\tau')=iG^0_{v';v'}(\tau)$ and $ G^0_{c';c'}(\tau')G^0_{c';c'}(\tau-\tau')=-iG^0_{c';c'}(\tau)$ in (\ref{Eqn:vvc_vertex}) and (\ref{Eqn:ccv_vertex}), respectively, which leads to

\begin{align}\label{Eqn:vvc_vertex_2}
    \Sigma_{1(im)}^{(3),(a'),-}(\tau)=  
   - i\!\!\!\!\!\!\!\!\! \sum_{\substack{vv'v''cc'\\v>v',v''>v'}} \!\!\!\!\!\!\!\!\!\!  v_{icv'v}\int_\tau^0 d\tau' G^0_{v;v}(\tau')G^0_{v';v'}(\tau)G^0_{c;c}(-\tau')
 v_{vc'v''c}G^0_{v''v''}(\tau-\tau') G^0_{c';c'}(\tau'-\tau)v_{v''v'c'm},
\end{align}
and 

\begin{align}\label{Eqn:ccv_vertex_2}
    \Sigma_{1(im)}^{(3),(a'),+}(\tau)= -i \!\!\!\!\!\!\!\!\! \sum_{\substack{cc'c''vv'\\c>c',c''>c'}} \!\!\!\!\!\!\!\!\!\!  v_{ivc'c}\int_\tau^0 d\tau' G^0_{c;c}(\tau')G^0_{c';c'}(\tau)G^0_{v;v}(-\tau')
 v_{cv'c''v}G^0_{c''c''}(\tau-\tau')G^0_{v';v'}(\tau'-\tau)v_{c''c'v'm}.
\end{align}

The other terms $(b')$, $(c')$, and $(d')$ are calculating changing the self-energy indices according to the diagrams

\begin{align}
    \Sigma^{(b')}_{1(im)}=\Sigma^{(a')}_{1(im)} &  &\text{with} & & \delta_{j'o''} v_{i'k''m''l'}\to -\delta_{j'm''} v_{i'k''o''l'} &  &\text{and} & & v_{i''j''l''m}\to - v_{i''j''ml''}  \nonumber \\
    \Sigma^{(c')}_{1(im)}=\Sigma^{(a')}_{1(im)} &  &\text{with} & & \delta_{j'o''} v_{i'k''m''l'}\to -\delta_{i'o''}v_{j'k''m''l'} &  &\text{and} & & v_{ik'o'm'}\to - v_{ik'm'o'} \nonumber \\
    \Sigma^{(d')}_{1(im)}=\Sigma^{(a')}_{1(im)} &  &\text{with} & &\delta_{j'o''} v_{i'k''m''l'}\to \delta_{i'm''} v_{j'k''o''l'}   &  &\text{and} & & v_{i''j''l''m}\to - v_{i''j''ml''}  \nonumber\\
    & &\text{and} &&v_{ik'o'm'}\to - v_{ik'm'o'}&&&&.
\end{align}
Using the same strategy as in Eqs (\ref{MCDE:vertex-1})-(\ref{Eqn:ccv_vertex_2}) these differences create
\begin{align}
    \Sigma^{(3),(b')}_{1(im)}=\Sigma^{(3),(a')}_{1(im)} &  &\text{with} & & v>v'' &  &\text{and} & & c>c'' \nonumber \\
    \Sigma^{(3),(c')}_{1(im)}=\Sigma^{(3),(a')}_{1(im)} &  &\text{with} & & v>v' &  &\text{and} & & c>c' \nonumber \\
    \Sigma^{(3),(d')}_{1(im)}=\Sigma^{(3),(a')}_{1(im)} &  &\text{with} & & v>v',v>v'' &  &\text{and} & &  c>c', c>c'',
\end{align}
By summing $\Sigma^{(3),(a')}_1$, $\Sigma^{(3),(b')}_1$, $\Sigma^{(3),(c')}_1$, and $\Sigma^{(3),(d')}_1$ and projecting in real space we arrive at (\ref{Diag:vertex}) with the \textit{restriction} $t\lessgtr t''\lessgtr t'$.

\end{widetext}

\end{document}